\documentclass[utf8]{style}

\usepackage{url,hyperref,lineno,microtype,cleveref}
\usepackage[onehalfspacing]{setspace}
\usepackage{array}
\usepackage{graphicx}
\usepackage{setspace}
\usepackage{multirow}
\usepackage{caption}
\usepackage{subcaption}

\def\keyFont{\fontsize{8}{11}\helveticabold }
\def\firstAuthorLast{Amadou {et~al.}} 
\def\Authors{Abdoul Aziz Amadou\,$^{1,2,*}$, Laura Peralta\,$^{1}$, Paul Dryburgh\,$^{1}$, Paul Klein\,$^{3}$, Kaloian Petkov\,$^{3}$, R. James Housden\,$^{1}$, Vivek Singh\,$^{3}$, Rui Liao\,$^{3\ddagger}$, Young-Ho Kim\,$^{3}$, Florin C. Ghesu\,$^{4}$, Tommaso Mansi\,$^{3\ddagger}$, Ronak Rajani\,$^{1}$, Alistair Young\,$^{1}$, and Kawal Rhode\,$^{1}$}
\def\Address{$^{1}$King’s College London, School of Biomedical Engineering \& Imaging Sciences, Department of Surgical \&
Interventional Engineering, London, United Kingdom\\
$^{2}$Siemens Healthcare Limited, Camberley, United Kingdom\\
$^{3}$Siemens Healthineers, Digital Technology and Innovation, Princeton, NJ, United States\\
$^{4}$Siemens Healthineers AG, Digital Technology and Innovation, Erlangen, Germany\\
$^{\ddagger}$Work done while at \textsuperscript{3}
}

\def\corrAuthor{Abdoul Aziz Amadou}
\def\corrEmail{abdoul.a.amadou@kcl.ac.uk}

\begin{document}
\onecolumn
\firstpage{1}

\title[Cardiac ultrasound simulation for navigation]{Cardiac ultrasound simulation for autonomous ultrasound navigation} 

\author[\firstAuthorLast ]{\Authors} 
\address{} 
\correspondance{}

\extraAuth{}

\maketitle

\begin{abstract}
Ultrasound is well-established as an imaging modality for diagnostic and interventional purposes. However, the image quality varies with operator skills as acquiring and interpreting ultrasound images requires extensive training due to the imaging artefacts, the range of acquisition parameters and the variability of patient anatomies. Automating the image acquisition task could improve acquisition reproducibility and quality but training such an algorithm requires large amounts of navigation data, not saved in routine examinations. Thus, we propose a method to generate large amounts of ultrasound images from other modalities and from arbitrary positions, such that this pipeline can later be used by learning algorithms for navigation.\\
We present a novel simulation pipeline which uses segmentations from other modalities, an optimized volumetric data representation and GPU-accelerated Monte Carlo path tracing to generate view-dependent and patient-specific ultrasound images.\\
We extensively validate the correctness of our pipeline with a phantom experiment, where structures’ sizes, contrast and speckle noise properties are assessed. Furthermore, we demonstrate its usability to train neural networks for navigation in an echocardiography view classification experiment by generating synthetic images from more than 1000 patients. Networks pre-trained with our simulations achieve significantly superior performance in settings where large real datasets are not available, especially for under-represented classes.\\
The proposed approach allows for fast and accurate patient-specific ultrasound image generation, and its usability for training networks for navigation-related tasks is demonstrated.

\tiny
\keyFont{ \section{Keywords:} Ultrasound, Monte Carlo Integration, Path Tracing, Simulation, Echocardiography}\\

\end{abstract}
\section{Introduction}
\label{sec:introduction}

\begin{figure*}[t]
    \centering
    \includegraphics[width=\textwidth]{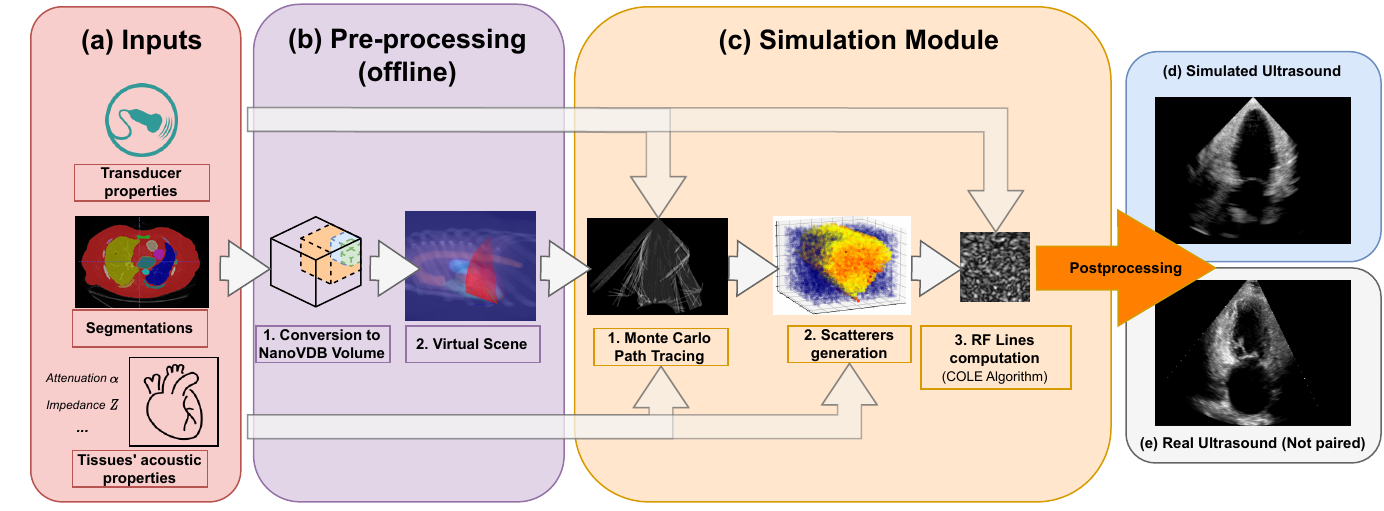}
    \caption{Simulation Pipeline. Using input segmentations from other modalities, transducer and tissue acoustic properties (a), we convert the segmentation to a NanoVDB volume (b.1) for ray tracing on the GPU. (b.2) shows a volume rendering of the ray tracing scene with various organs and the transducer's fan geometry. We model the sound waves as rays and perform ray tracing to simulate their propagation (c.1). We then generate a scattering volume (c.2) and compute the RF lines (c.3). Time-gain compensation and scan conversion are performed to yield the final simulation (d). A real ultrasound is shown for qualitative comparison (e).}
    \label{fig:sim_pipeline}
\end{figure*}

Ultrasound (US) is pivotal in the diagnosis, treatment and follow-up of patients in several medical specialities such as cardiology, obstetrics, gynaecology and hepatology. However, the quality of acquired images varies greatly depending on operators’ skills, which can impact diagnostic and interventional outcomes \cite{pinto2013}.

Providing guidance or automation for the image acquisition process would allow for reproducible imaging, increase both the workflow efficiency and throughput of echo departments and improve access to ultrasound examinations. This requires an intelligent system, capable of acquiring images by taking into consideration the high variability of patient anatomies.

Several works are investigating US acquisition automation but commercially available systems do not go beyond teleoperated ultrasound \cite{haxthausen2021}. Recent research towards autonomous navigation has used imitation learning \cite{huang2021} and deep reinforcement learning \cite{hase2020,li2023}. While these methods achieve varying degrees of success, they struggle to adapt to unseen anatomies, can only manage simple scanning patterns or are tested on small datasets.

The main advantage of a simulation environment is the ability to generate views that occur when operators navigate to a given standard view or anatomical landmark but are not saved in clinical routine. These datasets, which we call navigation data, can also contain imaging artefacts (e.g. shadowing caused by ribs). Hence, recent ultrasound image synthesis methods using neural networks \cite{gilbert2021,tiago2023} would struggle to generate these views as they require an understanding of ultrasound physics. Besides, learning-based approaches for navigation \cite{li2023} require a large number of images for training, including non-standard views, which are not available in classical ultrasound training datasets.

Using a simulation environment to train such a system would have several benefits. The trained model could learn while being exposed to a varying range of anatomies and image qualities, hence improving its robustness, and the training could be done safely, preventing the wear of mechanical components and potential injuries. This simulation environment should be: 1) Fast, to enable the use of state-of-the-art reinforcement learning algorithms. 2) Reproduce patients' anatomies with high fidelity. 3) Recreate attenuation artefacts such as shadowing. Moreover, exposing the system to a wide range of anatomies requires large-scale data generation capabilities, meaning the pre-processing of data must be streamlined.

This paper presents an ultrasound simulation pipeline using Graphical Processing Unit (GPU) based ray tracing on NVIDIA OptiX \cite{Parker2010}, capable of generating US images in less than a second. By combining networks capable of segmenting a wide range of tissues and a volumetric data representation, we overcome the scene modelling limitations of previous mesh-based simulation methods, enabling efficient processing of numerous datasets from different modalities.
Our pipeline, described in Fig. \ref{fig:sim_pipeline} takes as input segmentations of the organs of interest and, coupled with user-defined transducer and tissue properties, generates a simulated US by combining Monte Carlo path tracing (MCPT) and convolutional approaches.

Our contributions are the following:

\begin{itemize}
    \item Our pipeline is able to generate images from a large number of datasets from other modalities. Using an efficient GPU volumetric representation that allows for the modelling of arbitrary patient anatomies, and a Monte Carlo path tracing algorithm, we are able to synthesize more than 10,000 images per hour using a NVIDIA Quadro K5000 GPU. Furthermore, we demonstrate scalability by generating images from 1000 CT patient datasets in our experiments. In contrast, existing ray tracing methods limit their experiments to datasets one or two orders of magnitude smaller.

    \item We extensively validate the ability of our pipeline to preserve anatomical features through a phantom experiment by looking at distances and contrast between structures. Ultrasound image properties are further assessed by looking at first-order speckle statistics.
    
    \item We demonstrate the usability of our pipeline in training neural networks for transthoracic echocardiography (TTE) standard view classification, a task critical in ultrasound navigation guidance. The neural networks were initially pre-trained on synthetic images and subsequently fine-tuned using varying amounts of real data. With around half of the real samples, fine-tuned networks reach a performance level comparable to those trained with all the real data. We also report an improved classification performance when using pre-trained networks, particularly for under-represented classes.

\end{itemize}

This paper is organised in the following way: In Section 2.1, we provide an overview of relevant ultrasound simulation methods and highlight their limitations in terms of suitability as simulation environments. The next subsections in Section 2 detail our simulation implementation. Experimental results using a virtual phantom and a view classification network are shown in Section 3. This is followed by a discussion and a conclusion.
\section{Methods}

\subsection{Related Work}

Early methods were attempting to simulate the US image formation process by solving the wave equation using various strategies \cite{FieldIIa, FieldIIb,fullwave2009, karamalis2010,treeby2010}. While being accurate, these methods take a substantial amount of time to generate images (in the order of several minutes to hours \cite{karamalis2010,fullwave2009, FieldIIa, FieldIIb}), which is not scalable for large-scale training.

The COLE Algorithm developed by Gao et al.\cite{gao2009} is at the core of Convolutional Ray Tracing (CRT) methods. This approach allows for a fast simulation of ultrasound images with speckle noise by convolving a separable Point-Spread Function (PSF) with a scatterer distribution. Methods in \cite{burger2013,salehi2015,mattausch2018} replace the ray casting by ray tracing and combine it with the COLE algorithm to simulate images on the GPU. These methods follow a similar methodology where the input volumes are segmented and acoustic properties from the literature are assigned to each tissue. Scatterers amplitude are hyperparameters chosen such that the generated ultrasounds look plausible. Ray tracing is used to model large-scale effects at boundaries (reflection and refraction) and attenuation within tissue. Finally, the COLE algorithm is applied to yield the final image. The method developed in Mattausch et al. \cite{mattausch2018} distinguishes itself by employing MCPT to approximate the ray intensity at given points by taking into account contributions from multiple directions.

CRT methods enable fast simulations and the recreation of imaging artefacts. Methods in \cite{burger2013,mattausch2018} both make use of meshes to represent the boundaries between organs. However, using meshes comes with a set of issues as specific pre-processing and algorithms are needed to manage overlapping boundaries. This can lead to the erroneous rendering of tissues, hence limiting the type of scene that can be modelled, as reported in Mattausch et al.\cite{mattausch2018}. A further limitation of CRT methods lies in tissue parameterization, where scatterers belonging to the same tissue have similar properties, preventing the modelling of fine-tissue variations, and thus limiting the realism of the images.

Another line of work generates synthetic ultrasound images by directly sampling scatterers' intensities from template ultrasound images and using electromechanical models to apply cardiac motion \cite{alessandrini2015,alessandrini2018}. These are different from our line of work as they require pre-existing ultrasound recordings for a given patient, while we generate synthetic images from other modalities, which also enables us to simulate different types of organs other than the heart.

Finally, as deep learning has become increasingly popular, the field shifted towards the use of generative adversarial networks (GAN)  or diffusion models for image synthesis. These generative models have been used in several ways for image simulation: Either for generating images directly from segmentations \cite{gilbert2021,tiago2023,Stojanovski2023EchoFN}, calibrated coordinates \cite{hu2017}, or for improving the quality of images generated from CRT simulators \cite{vitale2019,Zhang2023UnpairedTF,tomar2021}. However, using GANs comes with several challenges: For instance, authors in Hu et al.\cite{hu2017} report mode collapse when generating images for poses where training data was not available and authors in Gilbert et al.\cite{gilbert2021} report hallucination of structures if anatomical structures are not equally represented in datasets. This suggests generative neural networks would struggle in generating out-of-distribution views or with image artefacts such as shadowing. This would be problematic for ultrasound navigation guidance as out-of-distribution views are frequently encountered before reaching a desired standard view.

Methods taking as input low-quality images from CRT simulators seem the most promising, but several works report issues in preventing the GANs from distorting the anatomy \cite{tomar2021} or introducing unrealistic image artefacts \cite{vitale2019}. While CRT methods are limited in realism, they match our requirements (speed, artefacts recreation, anatomical fidelity through accurate geometry) to train navigation/guidance algorithms.

\subsection{Pre-processing pipeline}

\begin{figure*}[t!]
    \centering
    \includegraphics[width=0.95\textwidth]{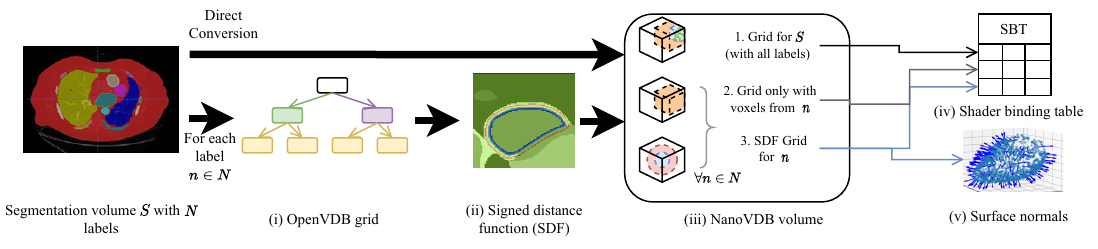}
    \caption{Overview of the pre-processing pipeline. A segmentation volume containing $N$ labels (one for each organ) is converted to a NanoVDB volume (iii) for use on the GPU. On the one hand, $S$ is directly converted to a grid containing all the labels (iii-1). On the other hand, for each label, an OpenVDB grid (i) containing only voxels belonging to the given label is created. In (ii), the SDF w.r.t the organ boundary is computed and used later during traversal to obtain surface normals (v). The final NanoVDB volume contains for each label, the corresponding voxel (iii-2) and SDF (iii-3) grids. Pointers to each grid are stored in the Shader Binding Table for access on the GPU (iv).}
    \label{fig:pre_processing_pipeline}
\end{figure*}

This section presents our novel pre-processing pipeline, shown in Figure \ref{fig:sim_pipeline}, which enables large-scale data generation by avoiding technical pitfalls caused by the use of meshes \cite{mattausch2018}, thus allowing us to model any anatomy. Besides, the use of segmentations is essential to implement constraints on the environment for navigation tasks.

Input volumes (Figure\ref{fig:sim_pipeline} (a)) are segmentations obtained from either CT or Magnetic Resonance Imaging (MRI) datasets, which are processed by a multi-organ segmentation algorithm inspired by Dong et al.\cite{dong2017}. The segmentation output contains all the structures relevant for echocardiography, e.g. individual ribs, sternum, heart chambers, aorta, and lungs.

During ray tracing, voxels need to be accessed at random. The access speed is highly dependent on the memory layout of the data. This problem has been addressed by OpenVDB \cite{museth2013} with its optimized B+ tree data structure and by its compacted, read-only and GPU-compatible version, NanoVDB \cite{museth2021}. Data in Open/NanoVDB are stored in grids. These grids can be written together into a single file, which we call an Open/NanoVDB volume. We convert the segmentation volumes into NanoVDB volumes (Figure\ref{fig:sim_pipeline} (b)) as described below.

A detailed overview of the pre-processing pipeline is shown in Figure \ref{fig:pre_processing_pipeline}. Firstly, the segmentation volume with all labels is converted to a NanoVDB grid (Figure \ref{fig:pre_processing_pipeline}, iii-1). This grid is used during ray tracing to access a label associated with a given voxel. Then, for each label in the segmentation volume, a narrow-band signed distance function (SDF) is computed such that the distance from voxels in the neighbourhood of the organ to its boundary is known (Figure \ref{fig:pre_processing_pipeline}, ii). Blue (resp. red) bands represent the voxels with negative (resp. positive) distance to that boundary, i.e. inside (resp. outside) it. The SDF grids are written to the output volume (Figure \ref{fig:pre_processing_pipeline}, iii-3) and are later used during traversal to compute smooth surface normals by looking at the SDF's gradient (Figure \ref{fig:pre_processing_pipeline}, v).

A separate grid containing only the voxels associated with the current organ is also saved (Figure \ref{fig:pre_processing_pipeline}, iii-2) in the output volume. Hence, the final NanoVDB volume (Figure \ref{fig:pre_processing_pipeline}, iii) contains the original voxel grid and, for each label, two grids: the SDF grid as well as the voxel grid. In practice, the pre-processing takes less than five minutes per volume and we use several worker processes to perform this task on multiple volumes in parallel.

\subsection{Scene Setup}

Similarly to previous work \cite{burger2013,salehi2015,mattausch2018}, the sound wave is modelled as a ray. The simulation is done using OptiX \cite{Parker2010}, which is a CUDA / C++ general-purpose ray tracing library providing its users with fast intersection primitives on the GPU.
The previously generated NanoVDB volume is loaded and the voxel grids corresponding to each label (Figure \ref{fig:pre_processing_pipeline}, iii-2)  are represented as Axis-Aligned Bounding Boxes (AABB) which are grouped together to create the Acceleration Structure (AS) used by OptiX to compute intersections.
We assign acoustic properties from the literature \cite{szabo2013} to each organ. A summary of all the assigned properties is listed in Table. \ref{table_tissue_props}. The values for $\mu_0, \mu_1, \sigma_0$ are the same as in Burger et al.\cite{burger2013}.
To retrieve data during traversal, OptiX uses a Shader Binding Table (SBT). We populate it with tissue properties, pointers to the organs' SDFs and a pointer to the original voxel grid (Figure \ref{fig:pre_processing_pipeline}, iv).
Finally, a virtual transducer is positioned in the scene. Transducer parameters are listed in Table. \ref{table_transducer_params}.

\begin{table}
\centering
\caption{List of properties assigned to tissues. Domain values are indicated for hyperparameters}
\begin{tabular}{|c|>{\centering\arraybackslash}p{8cm}|c|}
 \hline
 Property & Description & Domain \\ 
 \hline\hline
 Impedance ($Z$) & Tissue-specific acoustic impedance in $kg / (m^2 \cdot s)$ & -\\
 \hline
 Attenuation coefficient ($\alpha$) & Tissue-specific attenuation in $dB / (cm \cdot Hz) $ & - \\
 \hline
 Sound speed $(c)$ & Sound speed in a given tissue, in $m \cdot s^{-1}$ & - \\
 \hline
 $\mu_0, \sigma_0, \mu_1$ & Scatterer distribution parameter, from \cite{burger2013}. $\mu_0, \sigma_0$ control the scatterer amplitude while $\mu_1$ controls the probability of a scatterer being generated & $\mu_0, \sigma_0 \in [0, 1]$ \\
 \hline
  $\tau$ & Coefficient used to specify whether a reflection is more diffuse or specular, as in \cite{burger2013} & $\tau \in [0, 3]$\\
 \hline
  $\gamma$ &  Coefficient used to amplify small reflections, as in \cite{burger2013} & $\gamma \in [-2,2]$ \\
\hline
\end{tabular}
\label{table_tissue_props}
\end{table}

\subsection{Simulation Module}

The goal of the simulation module (Figure \ref{fig:sim_pipeline}(c)) is to generate view-dependent US images. This module is made of two parts.

The first part performs the ray tracing using OptiX. The goal of this module is to model large-scale effects (reflections, refractions and attenuation). This is done by computing, for each point along a scanline, the intensity $I$ sent back to the transducer.
The second part generates the US image by convolving the point spread function (PSF) with the scatterer distribution while taking into account the corresponding intensity $I(l)$ along the scanline.

\subsubsection{Background}

\paragraph{Ultrasound Physics:}

Here we first describe the phenomena happening during ray propagation: The wave loses energy due to attenuation following $I(l) = I_0 e^{-lf\alpha}$, with $I_0$ the initial wave intensity and $l$ the distance travelled in a given medium with attenuation $\alpha$ at frequency $f$.  When it reaches a boundary, it is partially reflected and transmitted depending on the difference in impedance between the two media. The reflection and transmission coefficients $R$ and $T$ are written:

\begin{align}
        	R(Z_1,  Z_2, \theta_1, \theta_2) &=  \Big( \frac{Z_2 cos(\theta_2) - Z_1 cos(\theta_1)}{Z_2 cos(\theta_2) + Z_1 cos(\theta_1)} \Big)^2 \\
	T(Z_1,  Z_2, \theta_1, \theta_2) &= 1 - R(Z_1,  Z_2, \theta_1, \theta_2) \\
	cos(\theta_1) &= \overrightarrow{n} \cdot \overrightarrow{v} \label{eq:cos_one}\\
	cos(\theta_2) &= \sqrt{1 - (\frac{Z_1}{Z_2})^2(1 - cos^2(\theta_1))} \label{eq:cos_two}
\end{align}

With $Z_1$ and $Z_2$ being the impedances of the media at the boundary, $\theta_1$ being the angle between the incident ray $\overrightarrow{v}$ and the surface normal $\overrightarrow{n}$ and $\theta_2$ the refracted angle.

\begin{table}
\centering
\caption{List of parameters used to configure the transducer}
\begin{tabular}{| c | >{\centering\arraybackslash} p{9cm}|} 
 \hline
 Property & Description \\ 
 \hline\hline
  Center Frequency & Transducer center frequency (in $Hz$) \\
 \hline
  Sampling Frequency & Signal sampling frequency (in $Hz$) \\
 \hline
 Element width & Width (in $mm$) of an element \\
 \hline
 Element height & Height (in $mm$) of an element \\
 \hline
 Kerf & Spacing between two elements (in $mm$) \\
 \hline
 Number of elements & Number of elements making up the matrix array \\
\hline
 Scan geometry & Type of scan geometry (e.g. linear, phased) \\
 \hline 
\end{tabular}
\label{table_transducer_params}
\end{table}

\paragraph{Rendering Equation:}
When the wave propagates in tissue, it can encounter several boundaries and bounce multiple times, depending on the scene geometry. Hence, retrieving the total intensity at a given point $P$ requires taking into account contributions coming from multiple directions. The field of computer graphics has faced similar challenges to compute global illumination.

We take inspiration from the rendering equation \cite{kajiya1986}:

\begin{equation} \label{eq:rendering_equation}
   L_{P \rightarrow \nu} = O_{P \rightarrow \nu} +  \int_{\Omega} f_{P, \omega \rightarrow \nu} L_{P \leftarrow \omega} cos(\theta) d\omega
\end{equation}

where:

\begin{itemize}
    \item $\Omega$ is the surface hemisphere around the surface normal at point $P$.
    \item $L_{P \rightarrow \nu}$ is the amount of light leaving point $P$ in direction $\nu$.
    \item $O_{P \rightarrow \nu}$ is the light emitted at $P$ in direction $\nu$.
    \item $f_{ P, \omega \rightarrow \nu}$ is a Bidirectional Scattering Distribution Function (BSDF) giving the amount of light sent back by a given material in direction $\nu$ when it receives light from direction $\omega$ at point $P$.
    \item $L_{P \leftarrow \omega}$ is the amount of light received by $P$ in direction $\omega$.
    \item Finally, $\theta$ is the angle between the surface normal at $P$, $\overrightarrow{n_P}$ and the incoming light direction $\omega$.
\end{itemize}

\subsubsection{Model derivation}

Several modifications are made to adapt Eq. \ref{eq:rendering_equation} to US physics. Firstly, the term $O_{P \rightarrow \nu}$ is zero in our case as scatterers do not emit echoes.

We can then refer to the intensity sent back to the transducer from $P$ as $I_{Tr}$. This term depends on the intensity $I(P)$ arriving at $P$, expressed as:

\begin{equation} \label{eq:accumulated_intensities}
	I(P) = \int_{\Omega}  I_{P' \rightarrow \omega} A_{P' \rightarrow P} d\omega
\end{equation}

This represents the accumulation of echoes reaching $P$ along directions $\omega$ from several points $P'$ located on other boundaries in the scene. This is illustrated in Figure \ref{fig:mc_logic} where contributions from $P_3$ and $P_2$ are gathered at $P$.  $I_{P' \rightarrow \omega}$ is the intensity leaving $P'$ in direction $\omega$ and $A_{P' \rightarrow P}$ is the attenuation affecting the wave from $P'$ to $P$ along $\omega$ (denoted as $\sim \alpha^-$ in Fig \ref{fig:mc_logic}). $I_{P' \rightarrow \omega}$ depends in turn on the intensity accumulated at $P'$ (illustrated by incident rays at $P_1 ... P_3$ in Figure \ref{fig:mc_logic}) following:

\begin{equation} \label{eq:intensity_and_brdf}
	I_{P' \rightarrow \omega} = I(P')f_{P', \omega' \rightarrow \omega} cos(\theta')
\end{equation}

With $\theta'$ the angle between the incident ray $\omega'$ \\ and $\overrightarrow{n_{P'}}$, and $f_{P', \omega' \rightarrow \omega} =  R(Z_1, Z_2, \omega', \omega)^{\beta} T(Z_1, Z_2, \omega', \omega)^{ 1 - \beta}$ where $\beta$ is a binary variable equal to one when the ray is reflected, and zero otherwise. We randomly choose whether to reflect or refract a ray and $\beta = 1$ when $u < R(Z_1, Z_2, \theta_1, \theta_2)$, with $u \sim U(0, 1)$, otherwise $\beta = 0$. Here $f$ is analogous to the BSDF in rendering and the corresponding loss of energy is represented at boundaries by $|-|$ in Figure \ref{fig:mc_logic}.

As we now have an expression for $I(P)$, we can compute $I_{Tr}$. This term depends on whether or not $P$ lies on an organ's surface. The two cases are described below:

\begin{itemize}
	\item Similarly to Burger \textit{et al.} \cite{burger2013}, on a boundary, the intensity reflected to the transducer $I_{Tr}(P) = I_R(P)$ is written as:

	\begin{equation} \label{eq:rendering_interfaces}
        	I_R (P ) = \Big( \frac{Z_2 - Z_1}{Z_2 + Z_1} \Big)^2  I(P)^\tau cos(\theta)^{\gamma}
	\end{equation}

	\item Otherwise, we simply have:
	\begin{equation} \label{eq:backscattering_interfaces}
        	I_{Tr}(P) = I(P)
	\end{equation}
\end{itemize}

The final signal, for a given point along a scanline with radial, lateral and elevation coordinates $(r,l,e)$, the received echo is formulated: 

\begin{equation} \label{eq:final_echo}
        E(r,l,e) = I_{Tr}(r,l,e)\rho(r,l,e) \otimes T(r,l,e)
\end{equation}

where $\rho(r,l,e)$ is a cosine modulated PSF and $T(r,l,e)$ the scatterer distribution.

\begin{align}
    \rho(x,y,z) &= exp\Big( -\frac{1}{2}\Big(\frac{r^2}{\sigma_r^2} + \frac{l^2}{\sigma_l^2} + \frac{e^2}{\sigma_e^2} \Big) \Big)cos(2\pi fr) \\
    T(r,l,e) &= \sum^N_{q=1} w_q a_q \delta(r - r_q)
\end{align}

$N$ is the number of scatterers, $a_q$ is the tissue-dependent scatterer amplitude, computed similarly to \cite{burger2013,salehi2015,mattausch2018}. Each scatterer is projected onto the scanline and associated with the closest radial sample $r_q$. Finally, $w_q$ is used to weight the contribution of a scatterer depending on its distance to the scanline. Let's write $\Delta_L$ and  $\Delta_E = e - e_S$ as the lateral and elevational distances of a scatterer to a scanline. Then $w_q$ can be computed in two ways:

\begin{itemize}
    \item Using an analytical beam profile, defined by a gaussian PSF with lateral and elevational variance $\sigma_L, \sigma_E$
    \begin{equation}
        w_q = exp \Big( -\frac{1}{2} \Big( \frac{\Delta_L^2}{\sigma_L^2} + \frac{\Delta_E^2}{\sigma_E^2} \Big) \Big) 
        \label{eq:gaussian_profile}
    \end{equation}
    \item Using a pulse echo field generated from Field II (offline) with the desired transducer configuration. The field is sampled based on $\Delta_L$ and $\Delta_E$ and the scatterer's radial depth. 
\end{itemize}

The computation of $E$ is done using the fast implementation of the COLE algorithm from Storve et al.\cite{storve2017}.

\subsubsection{Monte Carlo path tracing}

By substituting $I(P')$ in (\ref{eq:intensity_and_brdf}) by its expression in (\ref{eq:accumulated_intensities}), it is easy to see the recursive nature of the integral, which makes the problem hard to solve. Hence, we resort to Monte-Carlo integration, which is a useful tool to approximate high-dimensional integrals.

This allows us to write (\ref{eq:accumulated_intensities}) as:

\begin{equation} \label{eq:mc_formula}
        I(P) = \frac{1}{N} \sum_{i=1}^N  \frac{I(P_i) f_{P_i, \omega' \rightarrow \omega_i} A_{P_i \rightarrow P}  cos(\theta_i)}{p(\omega_i)}
\end{equation}

Unlike in Mattausch et al., \cite{mattausch2018}, we explicitly weight the pdf's contribution, $p(\omega_i)$, which is the probability of generating a sample in direction $\omega_i$. Indeed, at boundaries, rather than randomly varying the surface normal to choose a direction to trace reflected/refracted rays, we choose a random direction by sampling in a cone around the reflection/refraction directions, represented by the black arrow in Figure \ref{fig:subfigure_conic_pdf}. Indeed, when the wave hits large spherical scatterers, the reflected wavefront is a replica of the shape of the intersected area, which would take a conic shape in the case of spherical scatterers\cite{szabo2013}.

We generate random directions by sampling in spherical coordinates. More precisely, we have $\theta \sim U(0, 2\pi)$ and $\phi \sim \psi(\sigma, \mu, a ,b)$ where $\psi(\sigma, \mu, a ,b)$ is a truncated normal distribution. $\phi$ is sampled using inverse transform sampling. The joint distribution is $p(\theta, \phi) = \frac{\psi(\sigma, \mu, a,b)}{2\pi}$ and is illustrated in Figure \ref{fig:subfigure_conic_pdf}, where directions close to the reflection/refraction direction have a higher chance of being sampled (red colour) than the ones far from it (blue colour).

Finally, since we are working with solid angles, the distribution needs to be converted accordingly, with:

\begin{equation} \label{eq:joint_distribution}
    p(\omega) = \frac{p(\theta, \phi)}{sin(\theta)} = \frac{\psi(\mu, \sigma, a, b)}{2\pi sin(\theta)}
\end{equation}

When propagating, the sampled ray deviates from its main beam (blue, red and yellow rays in Figure \ref{fig:subfigure_mc_logic}, yielding a reduced echo intensity. Thus, in addition to the attenuation due to propagation through tissue, the sampled rays' intensities are further reduced by weighting them with a factor $w_R$ corresponding to the beam coherence (BC) as done in Mattausch et al \cite{mattausch2018}. For each point $P'$ along the sampled ray, the amplitude is weighted by $w_R = \frac{C_0}{C_0 + d(P, P')}$, where $C_0$ is a user-defined constant and $d(P, P')$ is the distance between $P'$ and its projection on the main beam $P$, as illustrated in Figure \ref{fig:subfigure_mc_logic}. We typically use $C_0$ values in the range $[0, 1]$.

\begin{subfigure}
    \begin{minipage}[t]{0.45\textwidth}
        \includegraphics[width=0.9\textwidth]{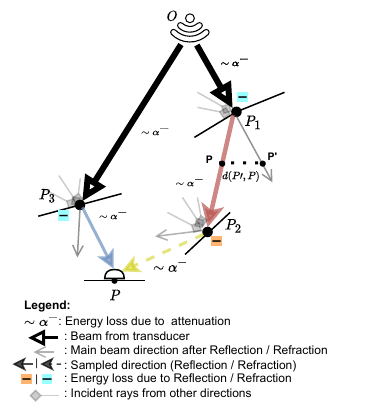}
    \end{minipage}%
    \centering
    \begin{minipage}[t]{0.45\textwidth}
        \includegraphics[width=0.75\textwidth]{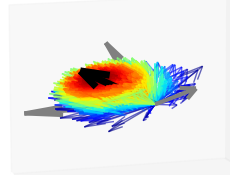}
    \end{minipage}
\par
    \setcounter{figure}{3}
    \setcounter{subfigure}{0}
    \begin{minipage}[b]{0.45\textwidth}
        \caption{Path tracing logic}
        \label{fig:subfigure_mc_logic}
    \end{minipage}%
    \setcounter{figure}{3}
    \setcounter{subfigure}{1}
    \begin{minipage}[b]{0.45\textwidth}
        \caption{Ray distribution at intersection}
        \label{fig:subfigure_conic_pdf}
    \end{minipage}

\setcounter{figure}{3}
\setcounter{subfigure}{-1}
    \caption{\textbf{(A)} A summary of the Monte Carlo path tracing logic: For a given point $P$ in the scene, we integrate the contributions from multiple waves reaching $P$ over its surface hemisphere. \textbf{(B)} A visualisation of the sampling pdf at intersections. The black arrow is analogous to the main beams in \textbf{(A)}. Directions close to the main beam (e.g. ray leaving $P_1$ in \textbf{(A)} have a higher chance of being sampled (thick red arrow) than the ones far from it (thick blue arrow, e.g. ray leaving $P_3$ in \textbf{(A)}.}
    \label{fig:mc_logic}
\end{subfigure}

\subsubsection{Traversal} Rays are sent from the virtual transducer depending on its scan geometry. The intersection with the volume is computed and from that point, we march stepwise along the ray using a hierarchical digital differential analyser (HDDA) \cite{museth2014}. At each step, the ray is attenuated and once a boundary is reached, we randomly reflect or refract the ray. We repeat the process until a maximum number of collisions is reached.

Once the RF scanlines are computed, we apply time-gain compensation, log compression, dynamic range adjustment and scan conversion to obtain the final simulated US.
\section{Experiments}

\setcounter{figure}{4}
\begin{subfigure}
\begin{center}
    \begin{minipage}[b]{0.26\textwidth}
        \includegraphics[width=\textwidth]{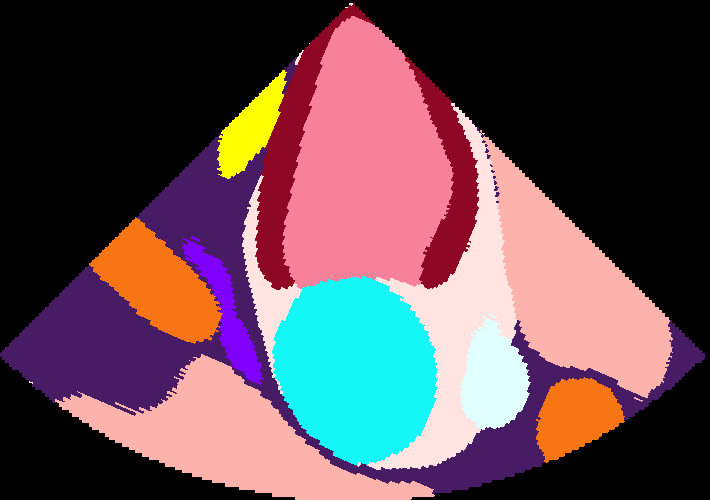}
        \caption{Segmentation}
        \label{fig:param_var_a}
    \end{minipage}
    \begin{minipage}[b]{0.26\textwidth}
        \includegraphics[width=\textwidth]{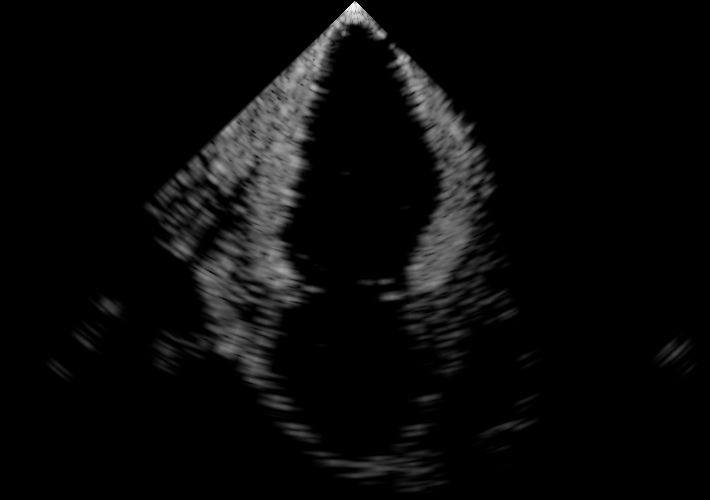}
        \caption{No MCPT}
        \label{fig:param_var_b}
    \end{minipage}
    \begin{minipage}[b]{0.26\textwidth}
        \includegraphics[width=\textwidth]{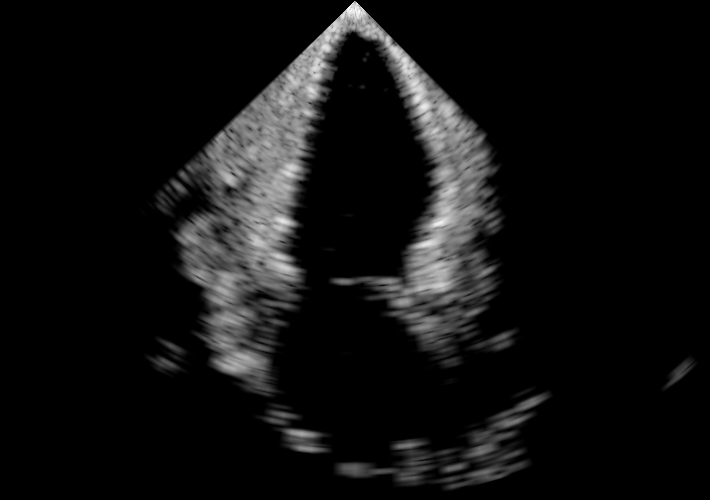}
        \caption{500 rays}
        \label{fig:param_var_c}
    \end{minipage}
\end{center}
    
\begin{center}
    \begin{minipage}[b]{0.26\textwidth}
        \includegraphics[width=\textwidth]{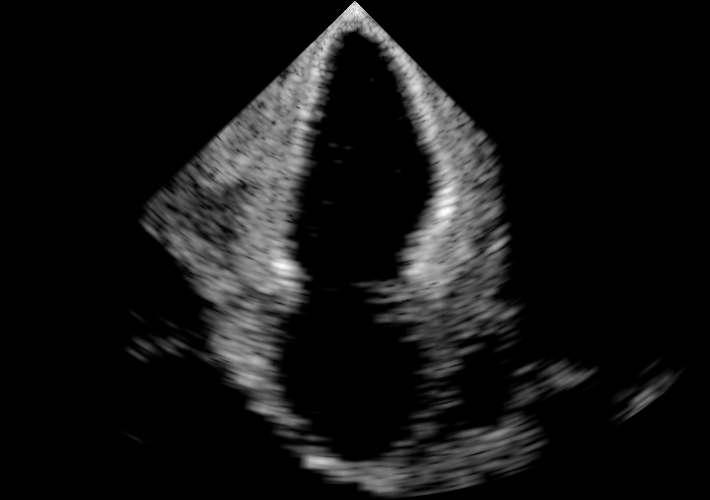}
        \caption{2500 rays}
        \label{fig:param_var_d}
    \end{minipage}
    \begin{minipage}[b]{0.26\textwidth}
        \includegraphics[width=\textwidth]{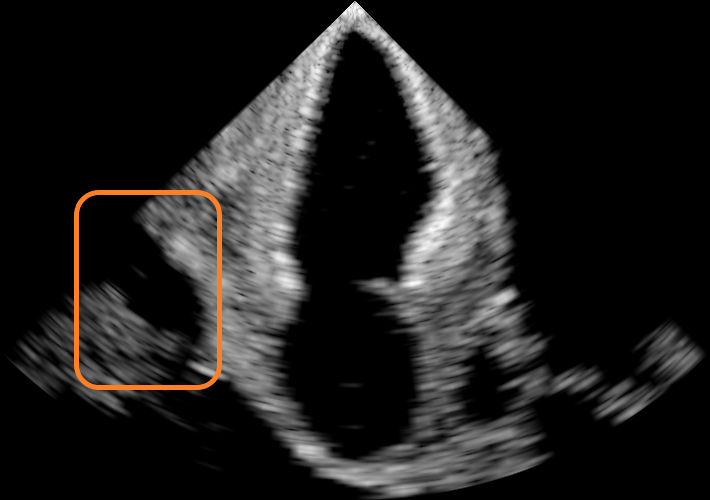}
        \caption{$C_0 = 0.2$ }
        \label{fig:param_var_e}
    \end{minipage}
    \begin{minipage}[b]{0.26\textwidth}
        \includegraphics[width=\textwidth]{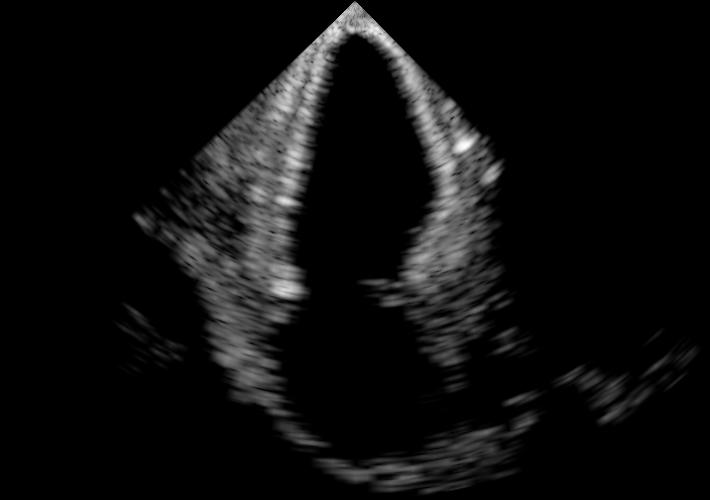}
        \caption{$\gamma=-1.8$ }
        \label{fig:param_var_f}
    \end{minipage}
\end{center}
\begin{center}   
    \begin{minipage}[b]{0.27\textwidth}
        \includegraphics[width=\textwidth]{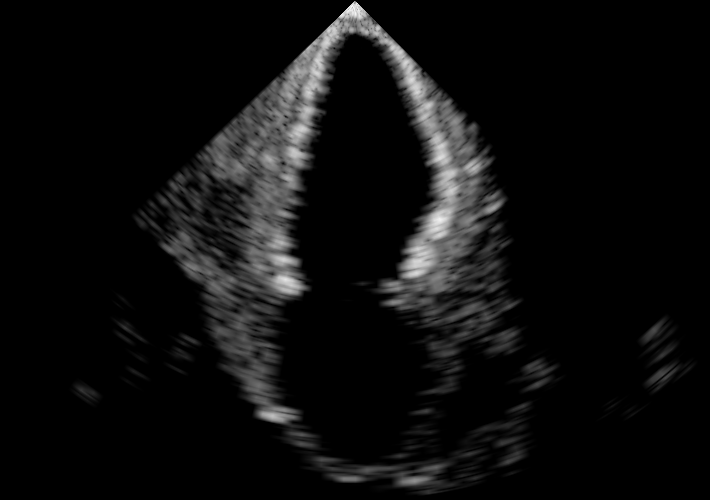}
        \caption{$\tau = 2.8$ }
        \label{fig:param_var_g}
    \end{minipage}
    \hspace{0.5cm}
    \begin{minipage}[b]{0.27\textwidth}
        \includegraphics[width=\textwidth]{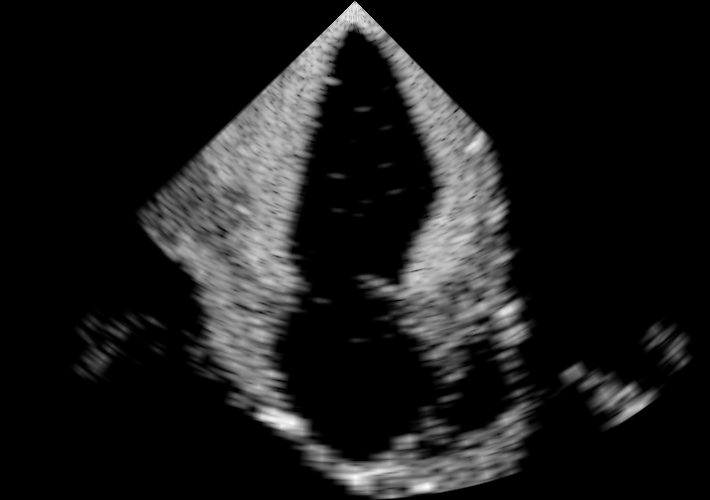}
        \caption{$\sigma_{L/E} = 1.5e^{-3}$ }
        \label{fig:param_var_h}
    \end{minipage}
\end{center}

\setcounter{figure}{4}
\setcounter{subfigure}{-1}
\caption{Illustration of the influence of the MCPT, beam coherence $C_0$ value, scatterer weighting strategy, $\tau$ and $\gamma$ terms. All simulations use MCPT, 2500 rays, a pulse-echo field from Field II with a focus at 60mm, $C_0 = 0.1$ and the myocardium properties are $\tau = 2.0$ and $\gamma = 0.1$ unless stated otherwise. \textbf{(A)} is an input segmentation map for an A2C view, where the orange label is associated with the aorta. In \textbf{(E)}, the orange box denotes the aorta, showing the simulations reproduce patient-specific anatomy with fidelity.}
\label{fig:param_variation}
    
\end{subfigure}

\begin{figure}[t!]
    \centering

    \includegraphics[width=0.5\textwidth]{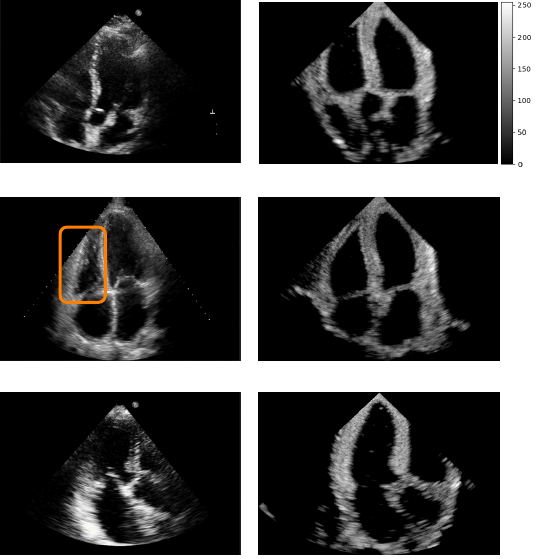}

    \caption{Real (left column) and simulated (right column) Apical 5, 4, 3 chambers views (top to bottom, not paired). The orange box denotes papillary muscles and fine cardiac structures which are not captured by the simulations, making the ventricles' borders sharper in the synthetic images.}
    \label{fig:qualitative_imgs}
\end{figure}

In the following sections, we begin by presenting qualitative results, where we examine the impact of different parameterizations and evaluate the pipeline's ability to replicate image artefacts and patient anatomies (Section 3.1).

Subsequently, we detail our phantom experiments, which serve as a validation of essential aspects of our simulation pipeline for its role as a learning environment. We assess its capability to reproduce anatomical structures by measuring physical distances and assessing contrast, using a calibration phantom as a reference. We further investigate its aptitude in generating a fully-formed speckle pattern, as speckle is an inherent property of ultrasound images (Section 3.2).

Lastly, we showcase the utility of these simulations in training a neural network for cardiac standard view classification, a critical task for ultrasound navigation guidance (Section 3.3).

\subsection{Qualitative Results}

Figure \ref{fig:param_variation} shows examples of simulated echocardiograms with various parameterizations: Firstly, the number of rays traced is critical in allowing the Monte Carlo process to converge and reveal the anatomy in the scene. Indeed, the left atrium is hardly visible in Figure \ref{fig:param_var_b} without MCPT, as rays reflect in deterministic directions, thus not propagating in the whole scene. Using MCPT with a greater number of rays improves the visibility of the anatomical structures as demonstrated in \Cref{fig:param_var_c,fig:param_var_d}. The beam coherence value $C_0$ impacts the intensity of the rays deviating from the main beams. This is illustrated in Figure \ref{fig:param_var_e} where a higher $C_0$ reveals the aorta as deviating rays are less attenuated. For our experiments and for future use as a training environment, the preferred simulation outcome would be similar to \Cref{fig:param_var_d,fig:param_var_e}, as critical structures for identifying the view are clearly visible.

\Cref{fig:param_var_f,fig:param_var_g} show an amplification of myocardium reflections in two ways using $\gamma$ and $\tau$: The reflection intensities in \Cref{fig:param_var_f} are angle-dependent while in \Cref{fig:param_var_g} all reflections are amplified. When using an analytical profile in Figure \ref{fig:param_var_h}, the axial distance of the scatterers along the scanline is not taken into account in $w_q$, meaning their amplitude is not attenuated with depth, yielding a brighter image in the far field.

\Cref{fig:qualitative_imgs} shows real acquisitions (left column) apical 5, 4, 3 chamber views (top to bottom) alongside simulations (right column). The chambers appear clearly in the images but the simulations lack fine tissue detail, as this information is lost when segmenting the input data. This is highlighted by the orange box in the four-chamber view, where the papillary muscles and valve leaflets in the real left ventricle acquisition make the ventricle's border fuzzier than in our simulation. Nevertheless, this shows the potential of the pipeline in generating any type of view.

\Cref{fig:artefacts} demonstrates post-acoustic enhancement and shadowing artefacts using a virtual sphere placed in a propagating medium. Post-acoustic enhancement is demonstrated in \Cref{fig:artefacts_a}, similar to artefacts caused by fluid-filled cystic structures in clinical settings. When using a highly reflective and attenuating sphere, a shadow is cast as in \Cref{fig:artefacts_b}. \Cref{fig:artefacts_c,fig:artefacts_d} illustrate acoustic shadowing in a more complex scene, where a rib is in front of the transducer. The advantage of our pipeline lies in its ability to produce such views, which are neither routinely saved nor available in open-source ultrasound datasets.

\begin{subfigure}

    \begin{center}
        \begin{minipage}[t]{0.3\textwidth}
            \includegraphics[width=\textwidth]{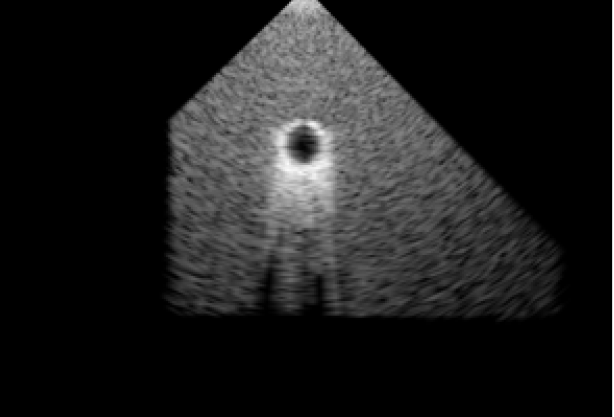}
            \captionsetup{format=hang, width=1.5\textwidth}
            \caption{Post-acoustic\\enhancement}
            \label{fig:artefacts_a}
        \end{minipage}%
        \hspace{0.5cm}
        \begin{minipage}[t]{0.3\textwidth}
            \includegraphics[width=\textwidth]{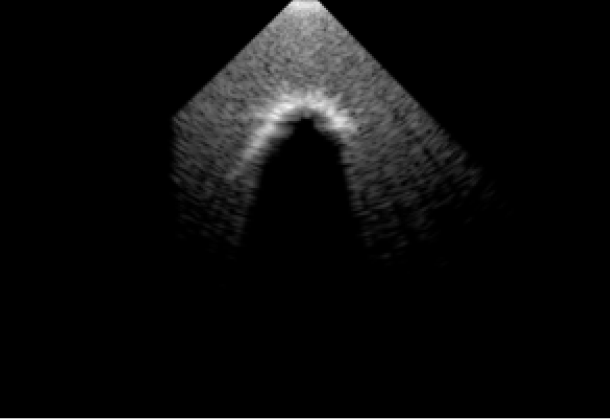}
            \caption{Acoustic shadowing}
            \label{fig:artefacts_b}
        \end{minipage}
    \end{center}
    
    \vspace{0.25cm} 
    \begin{center}   
        \begin{minipage}[b]{0.3\textwidth}
            \includegraphics[width=\textwidth]{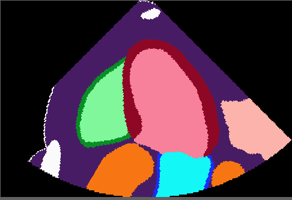}
            \captionsetup{format=hang, width=1.5\textwidth}
            \caption{Segmentation map}
            \label{fig:artefacts_c}
        \end{minipage}%
        \hspace{0.5cm}
        \begin{minipage}[b]{0.3\textwidth}
            \includegraphics[width=\textwidth]{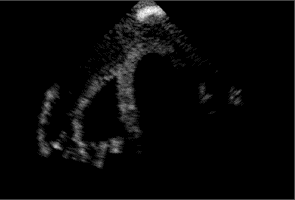}
            \caption{Rib shadowing}
            \label{fig:artefacts_d}
        \end{minipage}
    \end{center}

    \setcounter{figure}{6}
    \setcounter{subfigure}{-1}
    \caption{Our pipeline is able to recreate some artefacts such as \textbf{(A)} post-acoustic enhancement and \textbf{(B)} shadowing. Spheres filled with fluid \textbf{(A)} and with high attenuation \textbf{(B)} were used to recreate the artefacts. \textbf{(C)} shows segmentation labels of a scene with a rib in front of the transducer (white label) and \textbf{(D)} is the corresponding simulated image, demonstrating acoustic shadowing.}
    \label{fig:artefacts}
\end{subfigure}

\subsection{Phantom experiments}

We use a commercial calibration phantom (Multi-Purpose Multi-Tissue Ultrasound Phantom, model 040GSE, Sun Nuclear, USA) to perform the validation. Real acquisitions with a Siemens\textsuperscript{TM} Healthineers ACUSON P500\textsuperscript{TM} system (P4-2 phased transducer) are taken for lesion detectability comparison with the simulated images. To generate our simulations, a virtual phantom is built following the technical sheet describing the arrangement of structures in the phantom. Each type of structure is assigned a label and a segmentation volume is built. We simulate three different views, with each containing a different set of targets and perform various measurements on each synthesized view.  As we perform a comparison of lesion detectability in simulated and real images, we set the image pixel spacing of our simulations to the same value as the real acquisitions, i.e. at 0.23 mm. All simulations are done using a Desktop computer equipped with an NVIDIA Quadro K5000 GPU.

\setcounter{figure}{7}
\begin{subfigure}

    \begin{center}
        \begin{minipage}[t]{0.24\textwidth}
            \includegraphics[width=\textwidth]{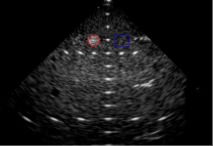}
            \caption{}
            \label{fig:gcnr_fig_a}
        \end{minipage}
        \begin{minipage}[t]{0.24\textwidth}
            \includegraphics[width=\textwidth]{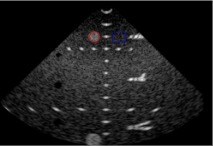}
            \caption{}
            \label{fig:gcnr_fig_b}
        \end{minipage}
        \begin{minipage}[t]{0.24\textwidth}
            \includegraphics[width=\textwidth]{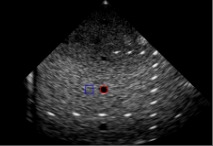}
            \caption{}
            \label{fig:gcnr_fig_c}
        \end{minipage}
        \begin{minipage}[t]{0.24\textwidth}
            \includegraphics[width=\textwidth]{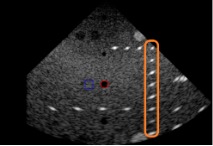}
            \caption{}
            \label{fig:gcnr_fig_d}
        \end{minipage}
    \end{center}

    \begin{center}
        \begin{minipage}[b]{0.24\textwidth}
            \includegraphics[width=\textwidth]{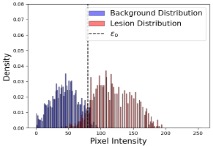}
            \caption{}
            \label{fig:gcnr_fig_e}
        \end{minipage}
        \begin{minipage}[b]{0.24\textwidth}
            \includegraphics[width=\textwidth]{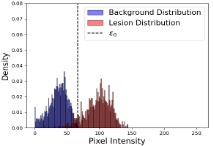}
            \caption{}
            \label{fig:gcnr_fig_f}
        \end{minipage}
        \begin{minipage}[b]{0.24\textwidth}
            \includegraphics[width=\textwidth]{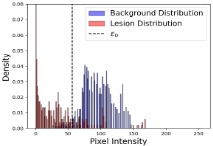}
            \caption{}
            \label{fig:gcnr_fig_g}
        \end{minipage}
        \begin{minipage}[b]{0.24\textwidth}
            \includegraphics[width=\textwidth]{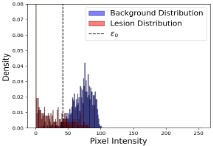}
            \caption{}
            \label{fig:gcnr_fig_h}
        \end{minipage}
    \end{center}

    \setcounter{subfigure}{-1}
    \caption{Examples of real and simulated views used in the lesion detectability and contrast experiment, alongside the corresponding histograms showing the lesion area and distribution (red) and the background area and distributions (blue). \textbf{(A,B)} Real and simulated acquisitions and the corresponding histograms (resp. \textbf{(E,F}) associated with the hyperechoic lesion. \textbf{(C,D)} Real and simulated acquisitions and the corresponding histograms (resp. \textbf{(G,H)}) associated with the anechoic lesion. In the histograms, $\epsilon_0$ denotes the optimal intensity threshold found that minimizes the probability of error when classifying pixels as belonging to the lesion or the background. The orange box in \textbf{(D)} denotes examples of targets used for the distance assessment.}
    \label{fig:gcnr_fig}
\end{subfigure}

\begin{table}
\centering
\caption{Parameters used in the experiments. For the view classification experiment, brackets indicate the range of values sampled.}
\begin{tabular}{|>{\centering\arraybackslash} p{6cm} |  c | c |}  
 \hline
 Property & \multicolumn{2}{|c|}{Value in experiment} \\ 
 \hline
 &  Phantom & View classification \\
 
 \hline \hline
\multicolumn{3}{|c|}{\textbf{Transducer parameters}} \\
 \hline
 Sampling frequency (MHz) & \multicolumn{2}{|c|}{50} \\
\hline
 Center frequency (MHz) & 3.6 & 3.6 \\
\hline
 Field II pulse echo field focus (cm) & 5 & 6.5 \\
 \hline
Analytical profile std ($\Delta_L, \Delta_E$) (mm) & N/A & 1.0 \\
 \hline
 
\multicolumn{3}{|c|}{\textbf{Simulation parameters}} \\
\hline
 Max num. collisions & 7 & 10 \\
 \hline
 Beam coherence $C_0$ & 0.1 & [0.01, 0.05, 0
 075, 0.1] \\
 \hline
 Num. rays per element & 1000 & [1000, 2000, 3000, 5000] \\
 \hline
 
\multicolumn{3}{|c|}{\textbf{Post-processing parameters}} \\
\hline
 Dynamic Range (dB) & 75 & [65, 75, 85, 95] \\
 \hline
 Time-Gain compensation (dB / cm) &  \multicolumn{2}{|c|}{1.5} \\
 \hline
 Reject threshold (dB) & 40 & [35, 45, 50, 60] \\
 \hline
 
\end{tabular}
\label{table:experiment_parameters}
\end{table}

\subsubsection{Experiment parameters}

The transducer, simulation and post-processing parameters for phantom and view classification experiments are listed in Table \ref{table:experiment_parameters}. For the phantom experiment, the transducer is parameterized similarly to the real one following the parameters listed in Table \ref{table_transducer_params}. The parameters for the truncated normal distribution $\phi$ are $\mu = 0, \sigma = \frac{\pi}{4}, a = 0, b = \frac{\pi}{2}$.

\subsubsection{Distance measurements}

We sampled pixels along a 1-D line going through nylon targets. The coordinates of the line were automatically computed given the technical phantom sheet. A 1-D signal was extracted from this line and peaks (corresponding to the centre of nylon wires) were identified.  Knowing the virtual transducer's position as well as the peaks' location along the line allowed us to compute a Target Registration Error (TRE) between the expected and simulated nylon wire positions. Examples of targets used in this experiment are shown by the orange box in Figure \ref{fig:gcnr_fig_d}. A detailed summary of the error per view and per target group is given in Table \ref{table:distance_measurements}. An error of $0.20 \pm 0.32$ mm was reported when measuring the TRE from 60 targets.

\begin{table}
\centering
\caption{Target Registration Error (TRE) between expected and simulated wire positions (mean $\pm$ std)}
\begin{tabular}{|>{\centering\arraybackslash} p{6cm} |>{\centering\arraybackslash} p{2cm} |>{\centering\arraybackslash} p{2cm} | >{\centering\arraybackslash} p{2cm} |}  
 \hline
 Structure & \multicolumn{3}{|c|}{TRE per view in mm} \\ 
\hline
 Target groups & View 1 & View 2 & View 3 \\
 \hline\hline
 Vertical Distance & $0.03 \pm 0.03$ & $0.09 \pm 0.07$ & $0.12 \pm 0.09$ \\ 
\hline
Horizontal Distance 1 (Near-Field) & $0.05 \pm 0.04$ & $0.16 \pm 0.14$ & $0.28 \pm 0.05$ \\
\hline
Horizontal Distance 2 (Far-Field) & $0.34 \pm 0.33$ & $0.25 \pm 0.59$ & $0.33 \pm 0.37$ \\
\hline
\end{tabular}
\label{table:distance_measurements}
\end{table}

A pattern emerges from Table \ref{table:distance_measurements}, where the error increases with depth (Horizontal Distance Groups 1 and 2). This is due to beam divergence in the far field, which decreases the lateral resolution. This agrees with experimental measurements.

\subsubsection{Lesion detectability and contrast}

Having an accurate contrast between background and surrounding structures is critical in ultrasound as it allows users to discriminate between tissues. Thus, we investigate the ability of our pipeline to simulate structures of various contrast. To this end, we compare anechoic and hyperechoic lesions from our virtual phantom to the same lesions from real acquisitions.

In addition to classical metrics such as Contrast to Noise Ratio (CNR) and contrast, we reported the generalized Contrast-to-Noise Ratio (gCNR)\cite{rodriguez2022}, a metric robust to dynamic range alterations and with a simple interpretation. Since our post-processing pipeline differs from the P500's as it is a commercial system, this metric would provide a way to compare the lesion detectability independently of post-processing differences.

We computed gCNR, CNR and contrast between lesions and background using two views. The background patch size was calculated to have a sample size similar to the lesion patch. Real and simulated acquisitions, as well as histograms of the lesions and background distributions, are illustrated in Figure \ref{fig:gcnr_fig}.

\begin{table}
\centering
\caption{gCNR, CNR and contrast (in dB), values from lesions in real and simulated US acquisitions (mean $\pm$ std). Simulations were generated 10 times to take in account the stochasticity of the MCPT and the scatterers' generation in $T(x)$}
\begin{tabular}{|p{4cm}|c|c|c|}
\hline
 Lesion & Metric & Real & Sim\\
 \hline \hline

 \multirow{3}{4cm}{Hyperechoic (+6 dB)} & gCNR &  0.19 & 0.22 $\pm$ 0.08 \\
 & CNR  & 0.27 & 0.29 $\pm$ 0.1 \\
 & Contrast & 4.72 & 4.88 $\pm$ 1.82 \\
 \hline
 \multirow{3}{4cm}{Hyperechoic (+15 dB)} & gCNR & 0.88 & 0.89 $\pm$ 0.02 \\
 & CNR & 0.74 & 0.77 $\pm$ 0.03 \\
 & Contrast & 16.07 & 16.97 $\pm$ 2.30 \\
 \hline
 \multirow{3}{4cm}{Anechoic 1} & gCNR & 0.80 & 0.82 $\pm$ 0.03 \\
 & CNR & 0.64 & 0.70 $\pm$ 0.05 \\
 & Contrast & -14.17 & -13.38 $\pm$ 1.89 \\
 \hline
 \multirow{3}{4cm}{Anechoic 2} & gCNR & 0.87 & 0.78 $\pm$ 0.04 \\
 & CNR & 0.72 & 0.70 $\pm$ 0.04 \\
 & Contrast & -16.61 & -13.38 $\pm$ 1.34 \\
 \hline
 \multirow{3}{4cm}{Anechoic 3} & gCNR &  0.71 & 0.71 $\pm$ 0.06 \\
 & CNR & 0.66 & 0.69 $\pm$ 0.06 \\
 & Contrast & -19.41 & -13.69 $\pm$ 2.43 \\
 \hline
\end{tabular}
\label{table:lesion_results}
\end{table}

A summary of the scores between real and simulated images is indicated in Table \ref{table:lesion_results}. Overall, gCNR, CNR and contrast values between real and simulated values are close, suggesting our pipeline reproduces lesions with fidelity. Contrast values for the second and third anechoic lesions differ as in the real acquisition, the far field is more hypoechoic compared to the focus area in the centre of the image.

\subsubsection{Speckle pattern analysis}

\begin{figure}[t]
\centering
\includegraphics[scale=0.3]{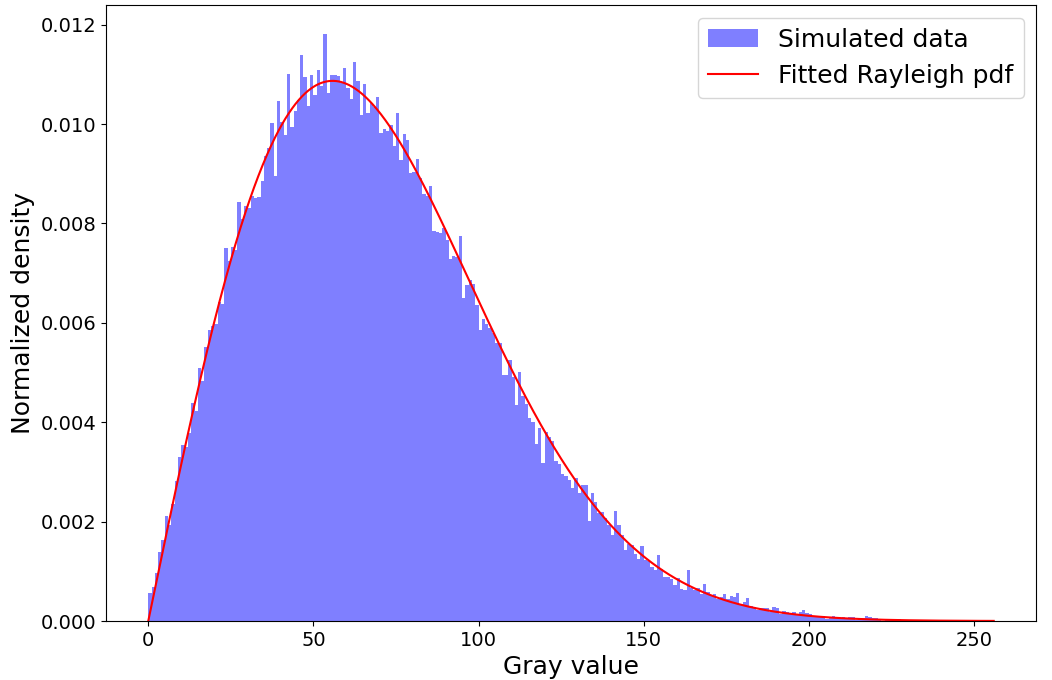}
\caption{Rayleigh distribution fit. The histogram shown is from a random run out of 10. We obtain a mean sum-of-squared Errors of $1.89e^{-5}$ w.r.t the fitted Rayleigh distribution and a SNR of $1.89 \pm 0.01$, which is in the ranges reported in the literature \cite{gao2009,gao2012,salehi2015}}
\label{fig:rayleigh_fit}
\end{figure}

In this section, we analyze the capability of our simulator to generate a fully-developed speckle pattern. To this end, we conduct a comparative analysis similar to Gao et al \cite{gao2012}, where random scatterers at a density of 600 $mm^{-2}$ and fixed amplitude of 1 are distributed in a 40 $\times$ 40 $mm^2$ volume. It is known for such an experiment that the envelope detected signal follows a Rayleigh distribution and its signal-to-noise ratio (SNR) reaches a value of 1.91 \cite{tuthill88}. The experiment is repeated 10 times to take into account its stochastic nature. Here, we use an analytical beam profile with $\Delta_E, \Delta_L = 2.0$ mm. For each run, the SNR is computed and the sum-of-squared errors (SSE) w.r.t a fitted Rayleigh distribution is calculated. An example histogram and fitted distribution from a run is shown in Figure \ref{fig:rayleigh_fit}.

We obtain a mean SSE of $1.81e^{-5}$ and SNR of $1.89 \pm 0.01$, which is in the ranges reported in the literature \cite{gao2009,gao2012,salehi2015}.  This suggests that our pipeline is able to create a fully developed speckle pattern.

\subsection{View classification}

Our last experiment assesses the usability of simulated images to train neural networks for view classification. This task is intrinsic to navigation as a network must be capable of identifying when a target view has been reached. We train networks to classify real apical views (A2C, A3C, A4C, A5C) and investigate the impact of fine-tuning with real data on the networks' performance, as networks trained in the simulation environment would likely be fine-tuned to adapt to real scenarios. Synthetic and real dataset generation are described in sections 3.3.1 and 3.3.2.
Furthermore, we conduct an ablation study where we evaluate the impact of parameters we believe impact the image quality the most, namely: the use of MCPT and the weighting method when projecting scatterers, i.e. with an analytical function or using a pulse echo field from Field II. The experimental setup is detailed in section 4.3.3, followed by the results in 4.3.4.

\subsubsection{Simulated TTE Dataset}

Chest CTs and Cardiac CTs from 1019 patients from the FUMPE \cite{fumpe} and The Cancer Imaging Archive \cite{tcia} (LIDC-IDRI \cite{lidc_idri}) datasets were used to generate simulated images. The volumes were automaticall segmented using \cite{dong2017} and pre-processed according to the pipeline described in Figure \ref{fig:pre_processing_pipeline} and several landmarks were automatically obtained (apex, the centre of the heart chambers...) and used to find the appropriate transducer orientations and positions to acquire the standard views.

For each view, we generate multiple synthetic samples by varying simulation parameters as described in Table \ref{table:experiment_parameters}. We generated more synthetic samples for the A5C view to compensate for the low number of datasets where we were able to automatically obtain a suitable view. The final dataset distribution is 30\%, 30\%, 30\% and 10\% resp. for the A2C, A3C, A4C and A5C classes.

All the samples from the simulated dataset are used for training. The average simulation time per image was 300 milliseconds. This number includes only the simulation step (i.e. Figure \ref{fig:sim_pipeline} (c)).

Finally, to conduct the ablation study, 3 different simulated datasets are created.

\begin{itemize}
    \item sim NO MCPT, where MCPT was disabled. Thus all samples are generated with deterministic raytracing.
    \item sim + MCPT, where MCPT was enabled and an analytical beam profile used.
    \item sim + MCPT + FIELD, where MCPT was enabled and a pulse echo field from Field II was used to weight the scatterers' contributions.
\end{itemize}

\subsubsection{Real TTE Dataset}

We used real US acquisitions to train and test the view classification network. The video sequences came from Siemens and Philips systems. During training, we sample randomly one frame from a given sequence and add it to the training batch. The real training dataset is also imbalanced, where the sample distribution in each fold for A2C, A3C, A4C and A5C classes is around 21\%, 18\%, 51\% and 10\%. 

\subsubsection{Evaluation methodology}
For this experiment, we used a Convolutional Neural Network (CNN) with a DenseNet architecture \cite{dense2017} to classify views. The network architecture is kept fixed for all experiments. Random weighted sampling is used to fight class imbalance. We divide the real dataset into 5 folds for cross-validation but always use the same synthetic dataset for pre-training.

In each fold, we create subsets $d_r$ of the real training dataset $D_{real}$ with varying amounts of real data. For each $d_r$, we train four networks: One network on $d_r$ only, to establish a baseline and then we pre-train 3 other networks on each one of the simulated datasets and fine-tune them on $d_r$. Validation and testing are always done on the same real datasets, independently of $d_r$'s size.

When pre-training, we employ the following data augmentations on the synthetic samples to match the variations observed in the real dataset: Cropping/zooming (e.g. to mimic real sequences where there's a zoom on a chamber or a valve), Gaussian smoothing, brightness and contrast jittering (to replicate varying texture qualities), fan angle variation (for real sequences where the fan angle is changed by the operator). No augmentations are applied to the real data.

When evaluating, for each video sequence, we perform a majority vote on the network's predictions on each frame to determine which label to assign to the sequence.

\begin{subfigure}  

    \begin{center}   
        \begin{minipage}[b]{0.35\textwidth}
            \includegraphics[width=\textwidth]{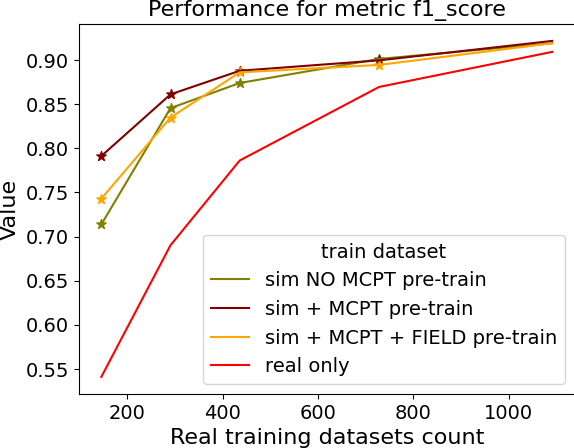}
            \caption{}
            \label{fig:view_classification_results_a}
        \end{minipage}
        \hspace{0.5cm}
        \begin{minipage}[b]{0.35\textwidth}
            \includegraphics[width=\textwidth]{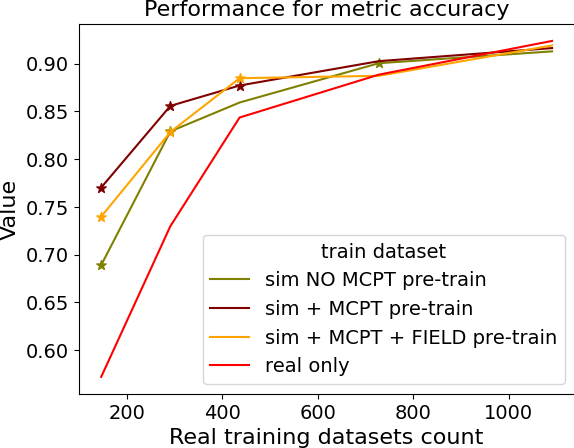}
            \caption{}
            \label{fig:view_classification_results_b}
        \end{minipage}
    \end{center}
    \begin{center}   
        \begin{minipage}[b]{0.35\textwidth}
            \includegraphics[width=\textwidth]{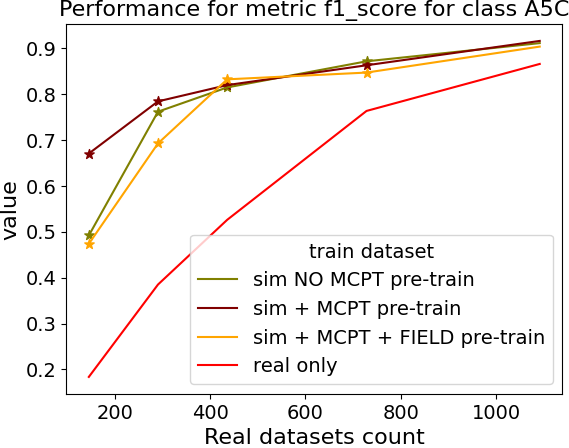}
            \caption{}
            \label{fig:view_classification_results_c}
        \end{minipage}
        \hspace{0.5cm}
        \begin{minipage}[b]{0.35\textwidth}
            \includegraphics[width=\textwidth]{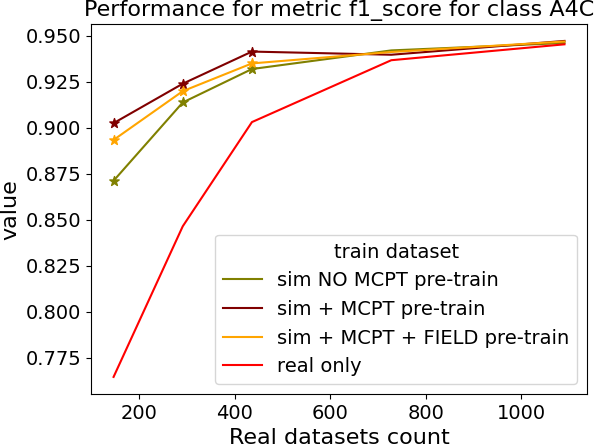}
            \caption{}
            \label{fig:view_classification_results_d}
        \end{minipage}
    \end{center}
    
    \setcounter{figure}{9}
    \setcounter{subfigure}{-1}
    \caption{Results of the view classification ablation study averaged over 5 folds. Networks pre-trained with simulations and then fine-tuned on real samples were compared to networks trained on real data only. The x-axis indicates the size of the subset of real data $d_r$. \textbf{(A,B)}report the F1-score and accuracy over the 4 classes while \textbf{(C,D)} report the metrics for the (most-represented) A4C and (under-represented) A5C classes. For a given $d_r$, a star is displayed on a graph if the p-value from a right-tailed Wilcoxon signed rank-test is $< 0.05$.}
    \label{fig:view_classification_results}
\end{subfigure}

\subsubsection{Results}

We report averaged F1-score and accuracy for all the classes in \Cref{fig:view_classification_results_a,fig:view_classification_results_b} and F1-score for the A5C and A4C classes in \Cref{fig:view_classification_results_c,fig:view_classification_results_d}.

Figure \ref{fig:view_classification_results} suggests pre-trained networks achieve a performance level comparable to networks trained on all real datasets when fine-tuned with at least half of the real data.

\begin{subfigure}
    \setcounter{figure}{10}
    \begin{center}   
        \begin{minipage}[b]{0.45\textwidth}
            \includegraphics[width=\textwidth]{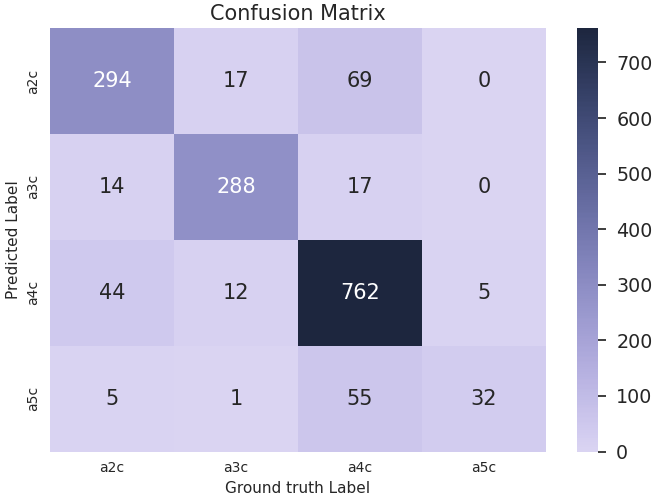}
            \caption{}
            \label{fig:confusion_matrices_a}
        \end{minipage}
        \hspace{0.5cm}
        \begin{minipage}[b]{0.45\textwidth}
            \includegraphics[width=\textwidth]{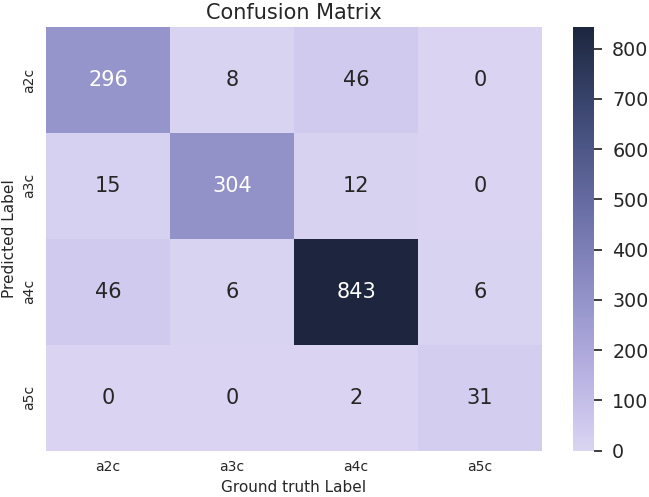}
            \caption{}
            \label{fig:confusion_matrices_b}
        \end{minipage}
    \end{center}

    \setcounter{figure}{10}
    \setcounter{subfigure}{-1}
    \caption{Confusion matrices for $d_r = 450$ in the view classification experiment. \textbf{(A)} Confusion matrix for the baseline trained on real data only. \textbf{(B)} Confusion matrix for the network pre-trained on simulated data with MCPT enabled. An analytical beam profile was used. The network pre-trained on simulated data \textbf{(B)} notably reduces the confusion between A5C and A4C classes.}
    \label{fig:confusion_matrices}

\end{subfigure}

Fine-tuned networks show significant improvements over their counterparts trained on real data (when $d_r < 800$). This trend is accentuated for the A5C class, which is the most under-represented in the dataset. Using simulated data for pre-training still benefits the dominant A4C class, as shown in \Cref{fig:view_classification_results_d}. Results for networks trained on simulated data only are not reported as they overfitted easily and performed poorly on the real test dataset.

Confusion matrices for $d_r = 450$ are reported in Figure \ref{fig:confusion_matrices} for the baseline trained on real data only (\Cref{fig:confusion_matrices_a}) and the network pre-trained on sim + MCPT (\Cref{fig:confusion_matrices_b}). There is a noticeable improvement in the results, highlighted by a reduction in confusion between the A5C and A4C classes.

Finally, no statistically significant differences were found when comparing the results of the networks pre-trained on sim +  MCPT and sim + MCPT + FIELD ($p > 0.05$). This suggests the choice of the weighting method for scatterers has little influence on neural network training on this task.
Results were statistically different between sim NO MCPT and sim + MCPT when $d_r < 450$ and only different between sim NO MCPT and sim + MCPT + FIELD when $d_r < 150$.
\section{Discussion}

In this section, we first discuss experimental results from the view classification experiment in 4.1. We then address the limitations of our proposed simulation pipeline in 4.2 and finish by expanding on potential applications of the pipeline and future work in 4.3.

\subsection{View classification}

In Figure \ref{fig:view_classification_results}, pre-trained networks show improved performance compared to the ones trained on real data only, meaning the simulations can be used to generate data when large datasets are not readily available or to target a sub-population which is less prevalent. This suggests our pipeline could be used to generate data for other tasks, given some improvements are made to circumvent the limitations caused by using segmentations. We expand on this in 4.2.

Moreover, networks pre-trained without MCPT achieved in some cases performances similar to their counterparts trained with MCPT. While MCPT allows for a better visibility of the anatomical structures as demonstrated in Figure \ref{fig:param_variation}, the discriminating features between views (i.e. heart chambers) are still present in the images without using MCPT. This would explain why the networks can still learn from such images. However, we believe using MCPT might be more critical in applications where all structures need to be clearly observable, such as image segmentation.

We limited the view classification experiment to four views as apical views were the only ones we could obtain robustly in an automatic way. Even then, we were not always successful in obtaining correct transducer orientations for each apical view in every patient dataset, especially for the A5C view. Indeed, view planes for each patient are obtained by finding landmarks using segmentations and morphological operations and then fitting a plane. Our automated method failed to consistently find a plane where the aorta and the four chambers were visible in the simulations.
This is related to the fact that we obtain our segmentations from CT data, where the patients are lying supine, and it is known that finding A5C views when patients are in the supine position is complicated in clinical settings as the imaging plane is suboptimal \cite{ugalde2018}. This explains the synthetic training data distribution in the view classification experiment. However, using an algorithm capable of navigating between views (which is what we intend to develop using the simulator), we could potentially generate datasets with a greater number of standard views.

Finally, in Figure \ref{fig:confusion_matrices}, there is a confusion between A2C and A4C classes. Our data is annotated such that all frames in a video sequence have the same label. However, there are multiple A4C sequences in which some frames resemble A2C views (due to suboptimal probe positioning or cardiac phase) but are labelled as A4C, which introduces confusion for the network during training.

\subsection{Proposed simulation pipeline}

While this pipeline allows for the fast simulation of arbitrary anatomies from a large number of patient datasets, it presents limitations:\\ 1) Similarly to other raytracing methods, we cannot simulate non-linear propagation. This prevents us from using techniques such as tissue harmonic imaging. Furthermore, we cannot reproduce reverberations. These could be simulated by summing the ray contributions temporally (i.e. by keeping track of the distance travelled by a ray) rather than spatially. However, this requires a careful weighting of the contribution of the randomly sampled rays with the beam coherence, so as to not yield incorrect results.\\
2) As seen in Figure \ref{fig:qualitative_imgs}, the border with the blood pool is sharp and the inhomogeneities of tissues are not represented in the simulations. This is due, respectively, to the segmentations, which do not capture details regarding smaller cardiac structures (e.g. papillary muscles, trabeculae ...) and to the assumption of homogeneity within the tissue (i.e. all scatterers' intensities in a given medium follow the same distribution) since the intensity variation between pixels is lost with the segmentations. While we could use the values from the input CT/MR volumes, the absence of a direct mapping between the scatterers' amplitudes and those values poses a challenge, making it difficult to circumvent this issue.

The impact of this limitation was illustrated in our attempt to train networks solely on simulated data for the view classification experiment, but the performance was poor. We noticed the network quickly overfitted the data. While the range of anatomies simulated is wide (+1000 patients), the lack of fine-tissue detail seems to limit the diversity of generated samples.
We believe a potential solution to this challenge would be a combination of our pipeline with generative models, to improve the realism and quality of simulations. This could enable the generation of large and realistic ultrasound datasets, with readily available anatomical labels.\\
3) While we do not address the topic of cardiac motion in this manuscript, it is possible to generate such sequences with our pipeline, given input volumes for each timestep of the cardiac cycle.\\
4) We recognize that the pressure applied by sonographers on the patient’s chest during TTE examinations can impact the image quality. We plan to address this in future work by incorporating a volumetric deformation model over the anatomical volume. Nevertheless, we note that the proposed framework would still be sufficient for training navigation algorithms for transesophageal imaging, where the impact on images of such anatomical shape deformations due to the ultrasound probe would be significantly smaller.

\subsection{Applications and future work}

We aim to use our pipeline as a simulator to train navigation algorithms, similar to Li et al \cite{li2023}. While the motivation behind the development of our pipeline is autonomous navigation, its capabilities could enable numerous downstream applications. Large dataset generation from segmentations could allow for the training of neural networks for tasks such as view classification, image segmentation or automated anatomical measurements.\\
In addition to the proposed use for automated acquisition, the method could be used for training or guidance of a semi-trained or novice ultrasound operator. Typically, guidance methods use 2D images from a pre-acquired 3D dataset. However, a simulation method would enable larger adjustments to the probe position.\\
While we focus on cardiac TTE imaging in this paper, other organs or modalities such as Transoesophageal Echocardiography (TEE) or Intracardiac Echocardiography (ICE), in 2D or 3D, could be simulated as a result of the built-in flexibility of our pipeline. Our future work will investigate both the use of the simulation pipeline as an environment to train deep reinforcement learning agents for autonomous navigation and the use of generative networks to improve the realism and train networks for several downstream tasks.
\section{Conclusion}

We have presented an ultrasound simulation pipeline capable of processing numerous patient datasets and generating patient-specific images in under half a second. In the first experiment, we assessed several properties of the simulated images (distances, contrast, speckle statistics) using a virtual calibration phantom. The geometry of our simulations is accurate, the contrast of different tissues is reproduced with fidelity and we are able to generate a fully developed speckle pattern.\\
We then synthesized cardiac views from more than 1000 real patient CT datasets and pre-trained networks using simulated datasets. The pre-trained networks required around half the real data for fine-tuning to reach a performance level comparable to networks trained with all the real samples, demonstrating the usefulness of simulations when large real datasets are not available.\\
The main limitation lies in the use of segmentations, unable to capture smaller cardiac structures or intensity variations between neighbouring pixels. Using a generative neural network to augment the simulations is a potential workaround. Such a pipeline enables a large number of downstream applications, ranging from data generation for neural network training (segmentation, classification, navigation) to sonographer training.

\section*{Conflict of Interest Statement}
The research was funded in part by Siemens Healthineers.

\section*{Disclosures}
The concepts and information presented in this paper/presentation are based on research results that are not commercially available. Future commercial availability cannot be guaranteed.

\section*{Author contributions}
AA: Implementation, Data curation, Methodology, Experiments, Writing, Visualization. LP: Methodology, Experiments. PD: Methodology, Experiments.  PK: Data curation. KP: Methodology. JH. Methodology. VS: Experiments, Writing. RL: Methodology, Supervision. YHK: Supervision, Writing. FG: Supervision, Writing. TM: Supervision, Resources. RR: Supervision, Writing. AY: Supervision, Writing. KR: Supervision, Writing, Resources

\section*{Funding}
This research was funded in part, by the Wellcome Trust under Grant WT203148/Z/16/Z and in part by Siemens Healthineers.

\section*{Acknowledgments}
The authors acknowledge the National Cancer Institute and the Foundation for the National Institutes of Health, and their critical role in the creation of the free publicly available LIDC/IDRI Database used in this study.\\
For the purpose of open access, the author has applied a CC BY public copyright licence to any Author Accepted Manuscript version arising from this submission.

\section*{Data Availability Statement}
The CT datasets presented in this article are publicly available \cite{lidc_idri,fumpe,tcia}.
The real ultrasound data cannot be shared publicly due to privacy reasons.
Part of the code used for the simulation is publicly available \cite{storve2017}.\\
For the purpose of open access, the author has applied a CC BY
public copyright licence to any Author Accepted Manuscript version
arising from this submission.

\bibliographystyle{style}
\bibliography{bibliography}

\begin{thebibliography}{39}
\expandafter\ifx\csname natexlab\endcsname\relax\def\natexlab#1{#1}\fi
\expandafter\ifx\csname urlstyle\endcsname\relax
  \expandafter\ifx\csname doi\endcsname\relax
  \def\doi#1{doi:\discretionary{}{}{}#1}\fi \else
  \expandafter\ifx\csname doi\endcsname\relax
  \def\doi{doi:\discretionary{}{}{}\begingroup \urlstyle{rm}\Url}\fi \fi
\expandafter\ifx\csname selectlanguage\endcsname\relax
  \def\selectlanguage#1{}\fi

\bibitem[{Pinto et~al.(2013)Pinto, Pinto, Faggian, Rubini, Caranci, Macarini
  et~al.}]{pinto2013}
Pinto AV, Pinto F, Faggian A, Rubini G, Caranci F, Macarini L, et~al.
\newblock Sources of error in emergency ultrasonography.
\newblock {\em Critical Ultrasound Journal\/} {\bf 5} (2013) S1 -- S1.

\bibitem[{Haxthausen et~al.(2021)Haxthausen, B{\"o}ttger, Wulff, Hagenah,
  Garc{\'i}a-V{\'a}zquez, and Ipsen}]{haxthausen2021}
Haxthausen FV, B{\"o}ttger S, Wulff D, Hagenah J, Garc{\'i}a-V{\'a}zquez V,
  Ipsen S.
\newblock Medical robotics for ultrasound imaging: Current systems and future
  trends.
\newblock {\em Current Robotics Reports\/} {\bf 2} (2021) 55 -- 71.

\bibitem[{Huang et~al.(2021)Huang, Xiao, Wang, Liu, Huang, and Sun}]{huang2021}
Huang Y, Xiao W, Wang C, Liu H, Huang RP, Sun Z.
\newblock Towards fully autonomous ultrasound scanning robot with imitation
  learning based on clinical protocols.
\newblock {\em IEEE Robotics and Automation Letters\/} {\bf 6} (2021)
  3671--3678.

\bibitem[{Hase et~al.(2020)Hase, Azampour, Tirindelli, Paschali, Simson,
  Fatemizadeh et~al.}]{hase2020}
Hase H, Azampour MF, Tirindelli M, Paschali M, Simson W, Fatemizadeh E, et~al.
\newblock Ultrasound-guided robotic navigation with deep reinforcement
  learning.
\newblock {\em 2020 IEEE/RSJ International Conference on Intelligent Robots and
  Systems (IROS)\/}  (2020) 5534--5541.

\bibitem[{Li et~al.(2023)Li, Li, Xu, Xiong, and Meng}]{li2023}
Li K, Li A, Xu Y, Xiong H, Meng MQH.
\newblock Rl-tee: Autonomous probe guidance for transesophageal
  echocardiography based on attention-augmented deep reinforcement learning.
\newblock {\em IEEE Transactions on Automation Science and Engineering\/}
  (2023).

\bibitem[{Gilbert et~al.(2021)Gilbert, Marciniak, Rodero, Lamata, Samset, and
  Mcleod}]{gilbert2021}
Gilbert A, Marciniak M, Rodero C, Lamata P, Samset E, Mcleod K.
\newblock Generating synthetic labeled data from existing anatomical models: An
  example with echocardiography segmentation.
\newblock {\em IEEE Transactions on Medical Imaging\/} {\bf 40} (2021)
  2783--2794.

\bibitem[{Tiago et~al.(2023)Tiago, Snare, Sprem, and Mcleod}]{tiago2023}
Tiago C, Snare SR, Sprem J, Mcleod K.
\newblock A domain translation framework with an adversarial denoising
  diffusion model to generate synthetic datasets of echocardiography images.
\newblock {\em IEEE Access\/} {\bf 11} (2023) 17594--17602.

\bibitem[{Parker et~al.(2010)Parker, Bigler, Dietrich, Friedrich, Hoberock,
  Luebke et~al.}]{Parker2010}
Parker SG, Bigler J, Dietrich A, Friedrich H, Hoberock J, Luebke DP, et~al.
\newblock Optix: a general purpose ray tracing engine.
\newblock {\em ACM SIGGRAPH 2010 papers\/}  (2010).

\bibitem[{Jensen et~al.(1992)Jensen, Svendsen, and Bruun}]{FieldIIa}
Jensen JA, Svendsen, Bruun NK.
\newblock Calculation of pressure fields from arbitrarily shaped, apodized, and
  excited ultrasound transducers.
\newblock {\em IEEE Transactions on Ultrasonics, Ferroelectrics and Frequency
  Control\/} {\bf 39} (1992) 262--267.

\bibitem[{Arendt(1996)}]{FieldIIb}
Arendt J.
\newblock Paper presented at the 10th nordic-baltic conference on biomedical
  imaging: Field: A program for simulating ultrasound systems (1996).

\bibitem[{Pinton et~al.(2009)Pinton, Dahl, Rosenzweig, and
  Trahey}]{fullwave2009}
Pinton GF, Dahl J, Rosenzweig SJ, Trahey GE.
\newblock A heterogeneous nonlinear attenuating full- wave model of ultrasound.
\newblock {\em IEEE Transactions on Ultrasonics, Ferroelectrics and Frequency
  Control\/} {\bf 56} (2009).

\bibitem[{Karamalis et~al.(2010)Karamalis, Wein, and Navab}]{karamalis2010}
Karamalis A, Wein W, Navab N.
\newblock Fast ultrasound image simulation using the westervelt equation.
\newblock {\em Medical image computing and computer-assisted intervention :
  MICCAI ... International Conference on Medical Image Computing and
  Computer-Assisted Intervention\/} {\bf 13 Pt 1} (2010) 243--50.

\bibitem[{Treeby and Cox(2010)}]{treeby2010}
Treeby BE, Cox BT.
\newblock k-wave: Matlab toolbox for the simulation and reconstruction of
  photoacoustic wave fields.
\newblock {\em Journal of biomedical optics\/} {\bf 15 2} (2010) 021314.

\bibitem[{Gao et~al.(2009)Gao, Choi, Claus, Boonen, Jaecques, van Lenthe
  et~al.}]{gao2009}
Gao H, Choi HF, Claus P, Boonen S, Jaecques SVN, van Lenthe GH, et~al.
\newblock A fast convolution-based methodology to simulate 2-dd/3-d cardiac
  ultrasound images.
\newblock {\em IEEE Transactions on Ultrasonics, Ferroelectrics and Frequency
  Control\/} {\bf 56} (2009) 404--409.

\bibitem[{B{\"u}rger et~al.(2013)B{\"u}rger, Bettinghausen, R{\"a}dle, and
  Hesser}]{burger2013}
B{\"u}rger B, Bettinghausen S, R{\"a}dle M, Hesser JW.
\newblock Real-time gpu-based ultrasound simulation using deformable mesh
  models.
\newblock {\em IEEE Transactions on Medical Imaging\/} {\bf 32} (2013)
  609--618.

\bibitem[{Salehi et~al.(2015)Salehi, Ahmadi, Prevost, Navab, and
  Wein}]{salehi2015}
Salehi M, Ahmadi SA, Prevost R, Navab N, Wein W.
\newblock Patient-specific 3d ultrasound simulation based on convolutional
  ray-tracing and appearance optimization.
\newblock {\em International Conference on Medical Image Computing and
  Computer-Assisted Intervention\/} (2015).

\bibitem[{Mattausch et~al.(2018)Mattausch, Makhinya, and
  Goksel}]{mattausch2018}
Mattausch O, Makhinya M, Goksel O.
\newblock Realistic ultrasound simulation of complex surface models using
  interactive monte‐carlo path tracing.
\newblock {\em Computer Graphics Forum\/} {\bf 37} (2018).

\bibitem[{Alessandrini et~al.(2015)Alessandrini, Craene, Bernard,
  Giffard‐Roisin, Allain, Waechter-Stehle et~al.}]{alessandrini2015}
Alessandrini M, Craene MD, Bernard O, Giffard‐Roisin S, Allain P,
  Waechter-Stehle I, et~al.
\newblock A pipeline for the generation of realistic 3d synthetic
  echocardiographic sequences: Methodology and open-access database.
\newblock {\em IEEE Transactions on Medical Imaging\/} {\bf 34} (2015)
  1436--1451.

\bibitem[{Alessandrini et~al.(2018)Alessandrini, Chakraborty, Heyde, Bernard,
  craene, Sermesant et~al.}]{alessandrini2018}
Alessandrini M, Chakraborty B, Heyde B, Bernard O, craene MD, Sermesant M,
  et~al.
\newblock Realistic vendor-specific synthetic ultrasound data for quality
  assurance of 2-d speckle tracking echocardiography: Simulation pipeline and
  open access database.
\newblock {\em IEEE Transactions on Ultrasonics, Ferroelectrics, and Frequency
  Control\/} {\bf 65} (2018) 411--422.

\bibitem[{Stojanovski et~al.(2023)Stojanovski, Hermida, Lamata, Beqiri, and
  G{\'o}mez}]{Stojanovski2023EchoFN}
Stojanovski D, Hermida U, Lamata P, Beqiri A, G{\'o}mez A.
\newblock Echo from noise: synthetic ultrasound image generation using
  diffusion models for real image segmentation.
\newblock {\em ASMUS@MICCAI\/} (2023).

\bibitem[{Hu et~al.(2017)Hu, Gibson, Lee, Xie, Barratt, Vercauteren
  et~al.}]{hu2017}
Hu Y, Gibson E, Lee LL, Xie W, Barratt DC, Vercauteren TKM, et~al.
\newblock Freehand ultrasound image simulation with spatially-conditioned
  generative adversarial networks.
\newblock {\em ArXiv\/} {\bf abs/1707.05392} (2017).

\bibitem[{Vitale et~al.(2020)Vitale, Orlando, Iarussi, and
  Larrabide}]{vitale2019}
Vitale S, Orlando JI, Iarussi E, Larrabide I.
\newblock Improving realism in patient-specific abdominal ultrasound simulation
  using cyclegans.
\newblock {\em International Journal of Computer Assisted Radiology and
  Surgery\/} {\bf 15} (2020) 183--192.

\bibitem[{Zhang et~al.(2023)Zhang, Portenier, and Goksel}]{Zhang2023UnpairedTF}
Zhang L, Portenier T, Goksel O.
\newblock Unpaired translation from semantic label maps to images by leveraging
  domain-specific simulations.
\newblock {\em ArXiv\/} {\bf abs/2302.10698} (2023).

\bibitem[{Tomar et~al.(2021)Tomar, Zhang, Portenier, and Goksel}]{tomar2021}
Tomar D, Zhang L, Portenier T, Goksel O.
\newblock Content-preserving unpaired translation from simulated to realistic
  ultrasound images.
\newblock {\em International Conference on Medical Image Computing and
  Computer-Assisted Intervention\/} (2021).

\bibitem[{Yang et~al.(2017)Yang, Xu, Zhou, Georgescu, Chen, Grbic
  et~al.}]{dong2017}
Yang D, Xu D, Zhou SK, Georgescu B, Chen M, Grbic S, et~al.
\newblock Automatic liver segmentation using an adversarial image-to-image
  network.
\newblock {\em ArXiv\/} {\bf abs/1707.08037} (2017).

\bibitem[{Museth(2013)}]{museth2013}
Museth K.
\newblock Vdb: High-resolution sparse volumes with dynamic topology.
\newblock {\em ACM Trans. Graph.\/} {\bf 32} (2013) 27:1--27:22.

\bibitem[{Museth(2021)}]{museth2021}
Museth K.
\newblock Nanovdb: A gpu-friendly and portable vdb data structure for real-time
  rendering and simulation.
\newblock {\em ACM SIGGRAPH 2021 Talks\/}  (2021).

\bibitem[{Szabo(2015)}]{szabo2013}
Szabo TL.
\newblock Diagnostic ultrasound imaging: Inside out (second edition).
\newblock {\em Ultrasound in Medicine and Biology\/} {\bf 41} (2015) 622.

\bibitem[{Kajiya(1986)}]{kajiya1986}
Kajiya JT.
\newblock The rendering equation.
\newblock {\em Proceedings of the 13th annual conference on Computer graphics
  and interactive techniques\/}  (1986).

\bibitem[{Storve and Torp(2017)}]{storve2017}
Storve S, Torp H.
\newblock Fast simulation of dynamic ultrasound images using the gpu.
\newblock {\em IEEE Transactions on Ultrasonics, Ferroelectrics, and Frequency
  Control\/} {\bf 64} (2017) 1465--1477.

\bibitem[{Museth(2014)}]{museth2014}
Museth K.
\newblock Hierarchical digital differential analyzer for efficient ray-marching
  in openvdb.
\newblock {\em ACM SIGGRAPH 2014 Talks\/}  (2014).

\bibitem[{Rodriguez-Molares et~al.(2019)Rodriguez-Molares, Rindal, D’hooge,
  M{\aa}s{\o}y, Austeng, Bell et~al.}]{rodriguez2022}
Rodriguez-Molares A, Rindal OMH, D’hooge J, M{\aa}s{\o}y SE, Austeng A, Bell
  MAL, et~al.
\newblock The generalized contrast-to-noise ratio: A formal definition for
  lesion detectability.
\newblock {\em IEEE Transactions on Ultrasonics, Ferroelectrics, and Frequency
  Control\/} {\bf 67} (2019) 745--759.

\bibitem[{Gao et~al.(2012)Gao, D’hooge, Hergum, and Torp}]{gao2012}
Gao H, D’hooge J, Hergum T, Torp H.
\newblock Comparison of the performance of different tools for fast simulation
  of ultrasound data.
\newblock {\em 2008 IEEE Ultrasonics Symposium\/}  (2012) 1318--1321.

\bibitem[{Tuthill et~al.(1988)Tuthill, Sperry, and Parker}]{tuthill88}
Tuthill TA, Sperry RH, Parker KJ.
\newblock Deviations from rayleigh statistics in ultrasonic speckle.
\newblock {\em Ultrasonic Imaging\/} {\bf 10} (1988) 81 -- 89.

\bibitem[{Masoudi et~al.(2018)Masoudi, Pourreza, Saadatmand-Tarzjan, Eftekhari,
  Zargar, and Rad}]{fumpe}
Masoudi M, Pourreza HR, Saadatmand-Tarzjan M, Eftekhari N, Zargar FS, Rad MP.
\newblock A new dataset of computed-tomography angiography images for
  computer-aided detection of pulmonary embolism.
\newblock {\em Scientific Data\/} {\bf 5} (2018).

\bibitem[{Clark et~al.(2013)Clark, Vendt, Smith, Freymann, Kirby, Koppel
  et~al.}]{tcia}
Clark KW, Vendt BA, Smith KE, Freymann JB, Kirby JS, Koppel P, et~al.
\newblock The cancer imaging archive (tcia): Maintaining and operating a public
  information repository.
\newblock {\em Journal of Digital Imaging\/} {\bf 26} (2013) 1045--1057.

\bibitem[{Armato and McNitt-Gray(2011)}]{lidc_idri}
Armato SG, McNitt-Gray MF.
\newblock The lung image database consortium (lidc) and image database resource
  initiative (idri): a completed reference database of lung nodules on ct
  scans.
\newblock {\em Medical physics\/} {\bf 38 2} (2011) 915--31.

\bibitem[{Huang et~al.(2016)Huang, Liu, and Weinberger}]{dense2017}
Huang G, Liu Z, Weinberger KQ.
\newblock Densely connected convolutional networks.
\newblock {\em 2017 IEEE Conference on Computer Vision and Pattern Recognition
  (CVPR)\/}  (2016) 2261--2269.

\bibitem[{Ugalde et~al.(2018)Ugalde, Medel, Romero, and Cornejo}]{ugalde2018}
Ugalde D, Medel JN, Romero C, Cornejo RA.
\newblock Transthoracic cardiac ultrasound in prone position: a technique
  variation description.
\newblock {\em Intensive Care Medicine\/} {\bf 44} (2018) 986--987.

\end{thebibliography}

\end{document}